\documentclass[fleqn,usenatbib]{mnras}

\usepackage{newtxtext,newtxmath}
\usepackage{amsmath}

\usepackage[T1]{fontenc}

\DeclareRobustCommand{\VAN}[3]{#2}
\let\VANthebibliography\thebibliography
\def\thebibliography{\DeclareRobustCommand{\VAN}[3]{##3}\VANthebibliography}

\usepackage{graphicx}	
\usepackage{amsmath}	
\definecolor{sienna}{RGB}{160,82,45}

\title[Irradiated disks parametric model]{A parametric model for externally irradiated protoplanetary disks with photoevaporative winds}

\author[L. Keyte \& T. J. Haworth]{
Luke Keyte$^{1}$\thanks{E-mail: l.keyte@qmul.ac.uk} and 
Thomas J. Haworth,$^{1}$
\\
$^{1}$Astronomy Unit, School of Physics and Astronomy, Queen Mary University of London, London E1 4NS, UK\\
}

\date{Accepted 2026 January 30. Received 2026 January 28; in original form 2025 December 10.}

\pubyear{\the\year{}}

\begin{document}
\label{firstpage}
\pagerange{\pageref{firstpage}--\pageref{lastpage}}
\maketitle

\begin{abstract}
Protoplanetary disks in massive star-forming regions may be exposed to ultraviolet radiation fields orders of magnitude stronger than the interstellar background. This intense radiation drives photoevaporative winds that fundamentally shape disk evolution and chemistry. However, full radiation hydrodynamic simulations of these systems remain computationally expensive, preventing systematic exploration of the parameter space. We present a parametric framework for efficiently generating density structures of externally irradiated protoplanetary disks with photoevaporative winds. Our approach implements a spherically diverging wind configuration with smooth transitions between the disk interior, the FUV-heated surface layer, and the wind itself. We validate this framework extensively against the \textsc{fried} grid of hydrodynamical simulations, demonstrating accurate reproduction of density structures across stellar masses from 0.3 to 3.0 $M_{\odot}$, disk radii from 20 to 150 au, and external FUV fields from $10^2$ to $10^5$ G$_0$. The complete framework is available as \textsc{puffin}, a Python package that generates full 1D or 2D density structures in seconds to minutes, compared to weeks or months for equivalent hydrodynamical calculations.  We demonstrate the scientific utility of this approach by modelling CO chemistry across a comprehensive parameter grid, using our density structures as inputs to thermochemical calculations. Our results show that external FUV irradiation significantly enhances CO gas-phase abundances through indirect heating mechanisms, which raise midplane temperatures and enhance thermal desorption of CO ice. This effect is strongest in the outer disk and scales with both external field strength and disk mass, with important implications for volatile budgets available to forming planets in clustered environments.
\end{abstract}



\section{Introduction}

Planet formation does not occur in isolation. Most stars, and by extension most planets, are born within dense stellar clusters where young protoplanetary disks are exposed to intense ultraviolet (UV) radiation from nearby massive stars \citep[e.g.][]{lada_lada_2003, kennicutt_evans_2012, krumholz_2019, winter_haworth_2022}. This external irradiation can exceed the interstellar background UV radiation field by factors of thousands or more, profoundly influencing disk evolution through photoevaporative mass loss \citep[e.g.][]{johnstone_1998, richling_yorke_2000, adams_2004, clarke_2007, facchini_2016, haworth_clarke_2019, concha-ramirez2019, ColemanAndHaworth22}. If we are to understand the diversity of planetary systems across the Galaxy, we must understand how these extreme radiation environments shape the birthplaces of planets.

Over the past few decades, the physical effects of external photoevaporation have been extensively characterised through both observations and theoretical modelling. Surveys of star forming regions, predominantly in the Orion Constellation, such as the Orion Nebula Cluster (ONC), NGC 1977, Sigma Ori, and NGC 2024 consistently show that irradiated disks tend to be more compact, less massive, and shorter-lived than disks in more quiescent environments \citep[e.g.][]{rigliaco_2009, mann_williams_2010, mann_2014, kim_2016, ansdell_2017, eisner_2018, van_terwisga_2020, haworth_2021, ballering_2023, mauco_2023, van_terwisga_2023, 2024A&A...687A..93A, aru_2024, mauco_2025, portilla_revelo_2025}. Complementary hydrodynamical simulations have provided insight into the mechanisms responsible for these changes, demonstrating that external far-ultraviolet (FUV) radiation can heat disk surface layers to several hundred Kelvin and drive photoevaporative winds from regions where the gas is only weakly gravitationally bound \citep{hollenbach_1994, johnstone_1998, storzer_hollenbach_1999, adams_2004, facchini_2016, haworth_2016}. Despite this progress in understanding the structural and dynamical evolution of irradiated disks, their chemical evolution under external irradiation remains comparatively unexplored.

The limited number of existing chemical studies of irradiated disks present a complex and sometimes contradictory picture, with outcomes that appear to depend sensitively on radiation field strength and disk region considered. Early models by \citet{walsh_2013} explored disks exposed to an intense external radiation field of $\sim 10^5$ G$_0$ (where G$_0 = 1.6 \times 10^{-3}$ erg cm$^{-2}$ s$^{-1}$ is the average interstellar radiation field; \citealt{habing_1968}). They found that although the inner disk remains largely shielded from external irradiation, the  outer disk can develop enhanced molecular abundances and stronger line emission. More recent work by \citet{calahan_2025} demonstrates that under certain conditions, elevated FUV fluxes can penetrate to the inner disk ($\lesssim 10$ au), reducing the radial extent of the H$_2$O snow surface and partially `resetting' midplane chemistry. Even moderate external FUV fields ($\lesssim 100$ G$_0$) can alter disk chemistry, as shown by \citet{gross_cleeves_2025}, who found that enhanced photodissociation and photodesorption in the surface layers and outer disk can boost the abundances of species such as C/C$^+$ and elevate diagnostic molecular ratios such as CN/HCN. However, 1D models that have coupled chemistry with disk evolution paint a different picture, predicting that the inner disk chemistry remains largely unchanged, retaining sub-solar C/O ratios and abundance gas-phase water \citep{ndugu_2024}.

Observational results are similarly mixed, though the interpretation is complicated by different observational techniques probing distinct disk regions. ALMA (Atacama Large Millimeter/submillimeter Array) observations, which trace the cooler outer disk and molecular emission, present contrasting findings. Surveys of two systems on the outskirts of the ONC find molecular line ratios and column densities indistinguishable from isolated disks \citep{diaz_berrios_2024}. Conversely, observations of other systems within the ONC reveal clear chemical differences, including near-interstellar CO abundances \citep{boyden_eisner_2023, goicoechea_2024} that contrast with the gas-phase carbon depletion inferred in many isolated disks \citep[e.g.][]{kama_2016b, maps_7_bosman_2021}. JWST spectroscopy, which probes the warmer inner disk, similarly yields mixed results. Observations of the irradiated T-Tauri disk XUE 1 reveal inner disk spectroscopic features consistent with non-irradiated counterparts \citep{ramirez_tannus_2023}, while the first observed externally irradiated Herbig disk exhibits distinctive chemistry, with strong CO$_2$ emission but a lack of water features typically seen in lower-mass systems \citep{frediani_2025}. These conflicting findings across different spatial scales expose critical gaps in our understanding and highlight the need for improved thermochemical models capable of identifying when and how external irradiation produces observable chemical signatures.

A major limitation of many existing chemical studies of irradiated disks is their reliance on simplified physical structures. Most adopt 1D slab models or standard power-law density profiles with isotropic background UV fields, and typically omit photoevaporative winds from the density structure altogether. Recent work, however, has demonstrated that these winds play an important role in shaping the chemistry of externally irradiated disks. \citet{keyte_haworth_2025} presented the first chemical models that explicitly incorporate the wind structure, revealing two major effects. First, although strong FUV radiation is quickly attenuated at or before the disk surface, the wind reprocesses this radiation into infrared emission that penetrates deep into the disk interior, heating the midplane and enhancing thermal desorption of molecular ices. Second, the wind itself contributes directly to atomic and molecular line emission, in some cases dominating the observed flux. These results indicate that neglecting the wind component can lead to fundamental misinterpretations of both the disk's internal composition and the origin of its emission. However, these models explored only a limited region of parameter space, leaving open the question of how these wind-driven effect vary across different stellar masses, disk properties, and external UV field strengths.

The main obstacle to systematic chemical studies across a broad parameter space is computational expense. Fully self-consistent radiation-hydrodynamic simulations, which are required to produce realistic disk-wind density structures for chemical modelling, can take weeks to months per model \citep[e.g.][]{haworth_clarke_2019}, making comprehensive parameter surveys infeasible. Yet such coverage is crucial for understanding how chemistry varies across different environments. What is needed is an efficient method to generate physically plausible disk-wind density structures that retain key physical features without the prohibitive cost of full simulations.

This paper presents a new parametric framework designed to meet this need. We develop and validate an efficient approach for constructing density structures of externally irradiated disks that include physically and empirically motivated photoevaporative winds, enabling systematic exploration of disk-wind chemistry across wide parameter ranges. Section \ref{sec:parametric_model} introduces the parametric model, beginning with a simplified one-dimensional formulation before extending to two dimensions. Section \ref{sec:chemical_modelling} applies this framework to generate a comprehensive grid of density structures spanning diverse stellar masses, disk masses and sizes, and external FUV field strengths, which are then used as inputs for thermochemical models. We focus on CO abundances as a case study to illustrate the scientific utility of our parametric approach. Section \ref{sec:discussion} discusses our findings in the context of recent observational and modelling studies, and Section \ref{sec:conclusions} summarises our conclusions.

\section{Parametric Model Framework}
\label{sec:parametric_model}

We present an approximate parametric model for determining the density structure of protoplanetary disks with external photoevaporative winds. While this model does not aim to replace full radiation hydrodynamic simulations, it provides an efficient tool for exploring the chemical and observational characteristics of disk-wind systems across a broad parameter space, enabling rapid generation of density structures that can serve as inputs for chemical models.

Our approach adopts a spherically diverging wind configuration, which we demonstrate accurately reproduces the behaviour of more sophisticated simulations under diverse physical conditions. Implementing this formulation requires determining the density at specific points within the flow. To develop intuition for this problem, we begin by examining the simplified one-dimensional case before extending the framework to two dimensions.

\subsection{1D model formulation}

One-dimensional models of external photoevaporation are built on the well-established principle that mass loss occurs predominantly at the outer edge of protoplanetary disks. This assumption is physically motivated: the outer disk region represents both the least gravitationally bound material and contains a substantial mass reservoir \citep{adams_2004, facchini_2016, haworth_2016}. These models solve the flow dynamics from the disk's outer edge and apply the resulting density profile over the solid angle subtended by this edge (up to one scale height) to determine mass-loss rate estimates.

This geometric framework assumes UV radiation propagates exclusively along the midplane toward the disk, while cooling radiation escapes in the opposite direction. All other directions are treated as infinitely optically thick. Despite this simplification, the 1D approach has proven remarkably effective, providing similar, but much less computationally expensive mass-loss rate estimates than analogous 2D models \citep{haworth_clarke_2019}. We adopt this established framework as the basis for our parametric implementation, which we describe in detail below.

\subsubsection{Model construction}

\begin{figure}
\centering
\includegraphics[clip=,width=1\linewidth]{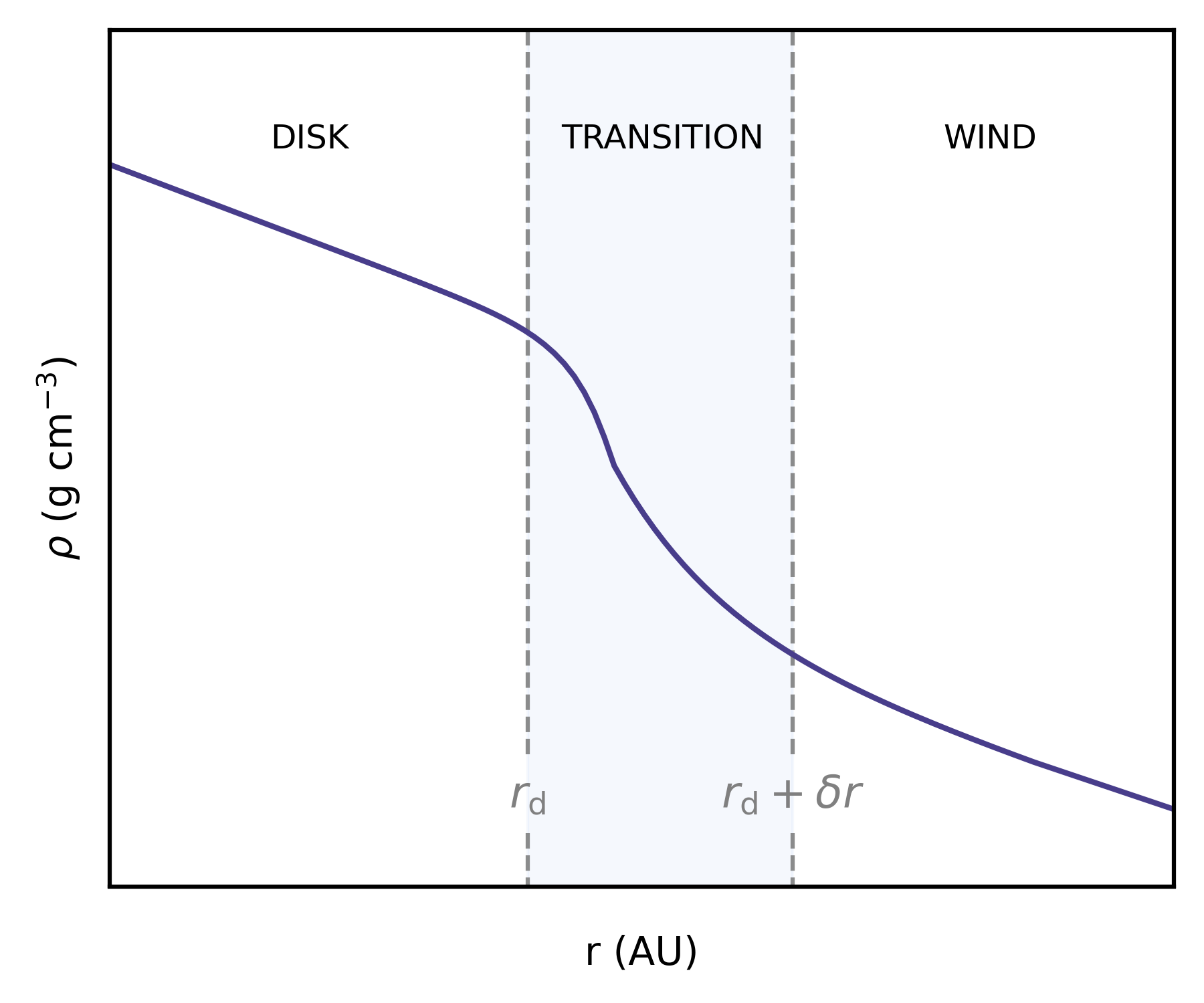}
\caption{Schematic illustration of the 1D parametric model density profile. The model comprises three distinct regions: (i) the disk interior with a power-law profile ($r=0$ to $r=r_\text{d}$), (ii) a smooth transition region where the exponential taper and `plateau' prescriptions bridge the disk and wind densities ($r=r_\text{d}$ to $r=r_\text{d} + \delta r$), and (iii) the outer photoevaporative wind following a spherically diverging profile ($r=r_\text{d} + \delta r$ to $r=r_\text{out}$). The vertical dashed lines mark the disk edge radius $r_\text{d}$, and the end of the transition region $\delta r$. The blue shaded area highlights the transition zone where disk and wind profiles are smoothly connected.}
\label{fig_1d_schematic}
\end{figure}

Our model domain extends from $r=0$ to $r=r_\text{out}$, with the central star at the origin. The density structure comprises three distinct components: the disk interior ($r < r_\text{d}$) which follows standard parametric formulations, a transition region ($r_\text{d}$ to $r_\text{d} + \delta r$) which smoothly connects the disk to the wind, and the outer photoevaporative wind ($r > r_\text{d} + \delta r$) which exhibits spherically diverging flow. This three-component structure, illustrated schematically in Figure \ref{fig_1d_schematic}, is designed to accurately capture the behaviour observed in hydrodynamical simulations.

We construct this structure by computing multiple density prescriptions and selecting the appropriate value at each location. For the disk interior, we compute a disk density at all radii that includes an exponential taper to capture the smooth decline beyond the nominal disk edge. The surface density follows:
\begin{equation}
    \Sigma(r) = \Sigma_{1\textrm{au}} \, \bigg(\frac{r}{\text{au}}\bigg)^{-1} \exp\bigg[-\bigg(\frac{r}{r_\text{d}}\bigg)^\gamma \bigg]
    \label{eq:sigma_tapered}
\end{equation}
where $\Sigma_{1\textrm{au}}$ is the surface density at 1 au, and the taper steepness parameter is:
\begin{equation}
    \gamma = A \; \bigg(\frac{M_*}{M_\odot}\bigg)^\alpha \bigg(\frac{r_\mathrm{d}}{\mathrm{au}}\bigg)^\beta \bigg(\frac{F_{\mathrm{FUV}}}{100 \;\mathrm{G}_0}\bigg)^{\varepsilon}
    \label{eq:gamma}
\end{equation}
where the normalisation constant $A$, and power-law indices $\alpha$, $\beta$, and $\varepsilon$ are empirical scaling parameters calibrated against hydrodynamical simulations (Section \ref{subsec:validating_1D}). This scaling ensures that more massive stars (which bind material more tightly) and stronger FUV fields produce steeper tapers at the disk edge.

In 1D, the disk temperature structure assumes a power-law dependence:
\begin{equation}
    T_\text{mid} = T_{1\textrm{au}}\, \left(\frac{r}{\textrm{au}}\right)^{-1/2}
\end{equation}
where $T_{1\textrm{au}}=150 \cdot (M_*/M_\odot)^{1/4}$ K is the midplane temperature at 1 au, following \citet{haworth_2018_fried}. The midplane density is derived from the surface density via:
\begin{equation}
    \rho_\text{disk}(r) = \frac{1}{\sqrt{2\pi}} \frac{\Sigma(r)}{h(r)}
    \label{eq:rho_disk}
\end{equation}
where $h=c_\text{s}/\Omega$ represents the scale height, with $c_\text{s}$ being the sound speed and $\Omega = (GM_*/r^3)^{1/2}$ the Keplerian angular velocity.

For radii $r \geq r_\text{d}$, we compute a spherically diverging wind density from the user-specified mass loss rate $\dot{M}$, following established prescriptions \citep[e.g.][]{adams_2004, winter_haworth_2022}:
\begin{equation}
   \rho_\text{wind}(r) = \frac{\dot{M}}{4\pi r^2 \mathcal{F} v_\text{R}} 
   \label{eq:rho_wind}
\end{equation}
where $\mathcal{F}=h_\text{d}/\sqrt{h_\text{d}^2+r_\text{d}^2}$ represents the solid angle fraction subtended by the outer disk edge, and $v_\text{R}$ denotes the flow velocity. To determine the flow velocity, we analysed 1D external photoevaporation models from the \textsc{fried} grid \citep{haworth_2018_fried, haworth_2023_fried2}, specifically utilizing the fiducial microphysics subset from the second version of the grid (dust-depleted wind and a PAH-to-dust ratio typical of that in the ISM, $f_\text{PAH}=1$). We approximate $v_\text{R}$ as equal to the local sound speed in the wind:
\begin{equation}
   c_\text{s,wind} = \sqrt{\frac{k_\text{B}T_\text{PDR}}{\mu_\text{wind} m_\text{p}}} 
   \label{eq:c_s_wind}
\end{equation}
where the photodissociation region temperature scales with the external FUV field:
\begin{equation}
    T_\text{PDR} = T_0 \, \bigg(\frac{F_\text{FUV}}{1000 \;\text{G}_0} \bigg)^{0.2}
    \label{eq:t_pdr}
\end{equation}
with $T_0 = 200$~K being the reference temperature at $1000$ G$_0$. This parameterisation represents the temperature at the wind base. The shallow scaling exponent reflects grain charging effects that reduce photoelectric heating efficiency at higher FUV fields \citep{tielens_hollenbach_1985}, producing a sub-linear temperature dependence on the incident flux. We impose minimum and maximum temperature limits of 10~K and 3000~K, respectively.

For weaker external FUV fields ($\lesssim 1000$ G$_0$), hydrodynamical models show an extended `plateau' region beyond $r_\text{d}$ where the density in the transition region decreases more gradually than produced by the exponential taper alone. Rather than dropping sharply to the wind density, the transition region maintains intermediate densities over a substantial radial extent. To capture this behaviour across the full range of irradiation conditions, we define an additional density prescription for $r > r_\text{d}$ that smoothly interpolates between the disk density at $r_\text{d}$ and the wind density at larger radii:
\begin{equation}
    \rho_\text{plateau}(r) = \rho_{\text{disk}}(r_\text{d}) \left( \frac{\rho_{\text{wind}}(r)}{\rho_{\text{disk}}(r_\text{d})} \right)^{f(r)}
    \label{eq:rho_plateau}
\end{equation}
The blending function $f(r)$ varies smoothly from 0 to 1 in logarithmic space to handle the many orders of magnitude variation in density:
\begin{equation}
    f(r) = \frac{1 - \exp\left[-\lambda \, \log_{10}(r/r_\text{d})\right]}{1 - \exp(-\lambda)}
    \label{eq:blending}
\end{equation}
The parameter $\lambda$ controls how rapidly the transition occurs from plateau to wind:
\begin{equation}
    \lambda = r_\text{d}^{p} + \bigg(\frac{F_\text{FUV}}{1000 \; \text{G}_0}\bigg)^{q}
    \label{eq:k_param}
\end{equation}
where $p$ and $q$ are fitted constants (Section \ref{subsec:validating_1D}). At weak fields ($F_\text{FUV} \sim 100$ G$_0$), $\lambda$ is small, producing a very gradual transition. At strong fields ($F_\text{FUV} \sim 10^5$ G$_0$), $\lambda$ is large, creating a sharp drop that approaches a direct disk-to-wind transition.

The final density at each radius is determined by selecting the maximum value from all applicable prescriptions:
\begin{equation}
    \rho(r) = \begin{cases}
        \rho_\text{disk}(r) & \text{if } r < r_\text{d} \\
        \max[\rho_\text{disk}(r), \, \rho_\text{wind}(r), \, \rho_\text{plateau}(r)] & \text{if } r \geq r_\text{d}
    \end{cases}
    \label{eq:rho_final}
\end{equation}
This formulation ensures physically consistent behaviour across all radii. In the inner regions ($r < r_\text{d}$), the disk density dominates. Beyond the disk edge, the exponential taper in $\rho_\text{disk}$ provides a smooth decline, while $\rho_\text{plateau}$ captures the extended transition at weak FUV fields, and $\rho_\text{wind}$ governs the outer spherically diverging flow. The ``max'' operation in Equation \ref{eq:rho_final} allows the physically relevant prescription to naturally emerge at each radius depending on the local conditions.

\begin{figure}
\centering
\includegraphics[clip=,width=1\linewidth]{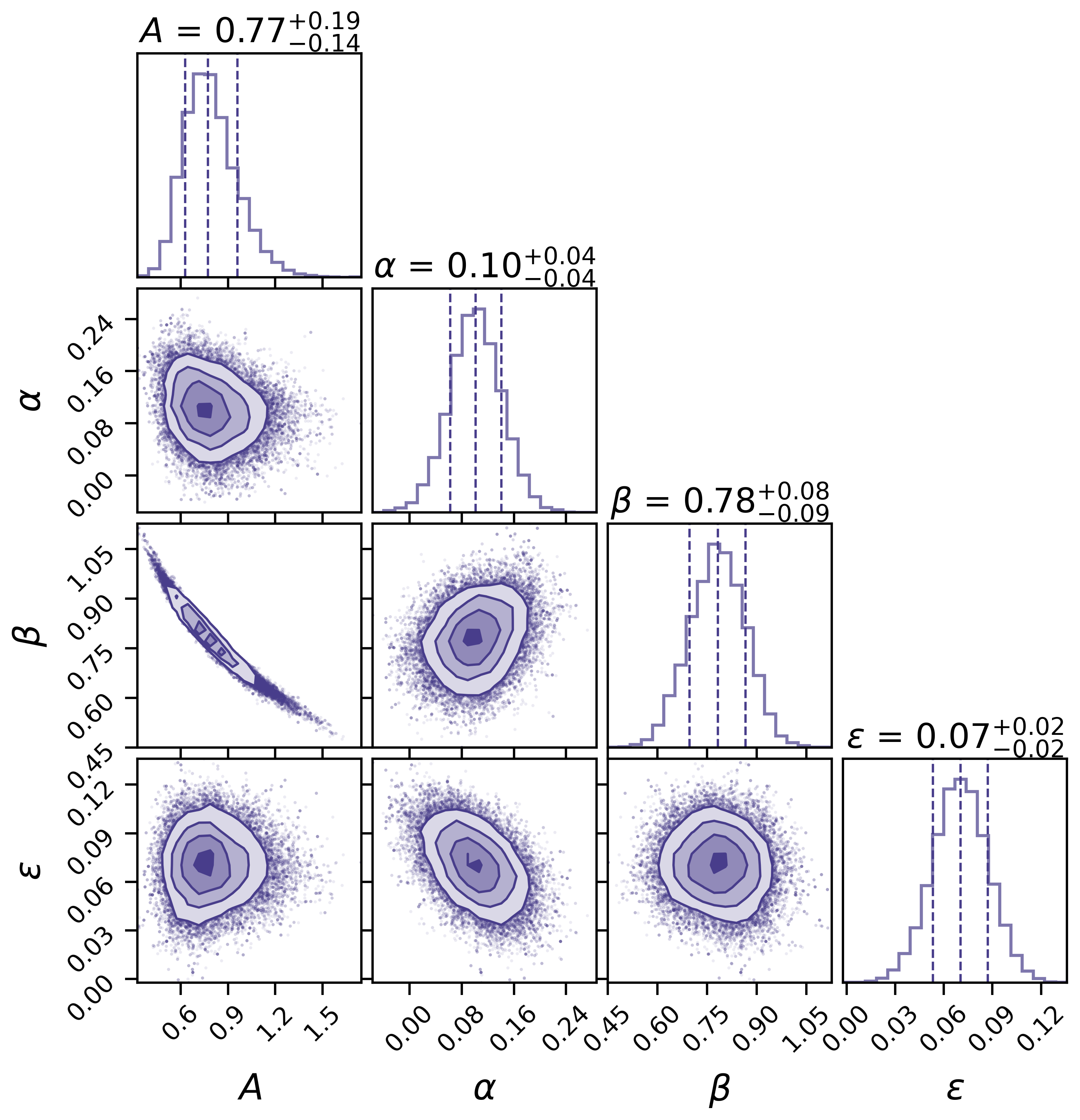}
\caption{Posterior distributions for the parameters governing the exponential taper that links the disk outer edge to the spherically expanding wind in our 1D model. The parameter $A$ sets the overall normalisation, while $\alpha$, $\beta$, and $\varepsilon$ describe the dependence of the taper on stellar mass, disk radius, and external FUV field strength, respectively. The MCMC was run for 40{,}000 iterations and yields well-converged constraints on all four parameters.}
\label{fig_corner_plot}
\end{figure}

\begin{figure*}
\centering
\includegraphics[clip=,width=1\linewidth]{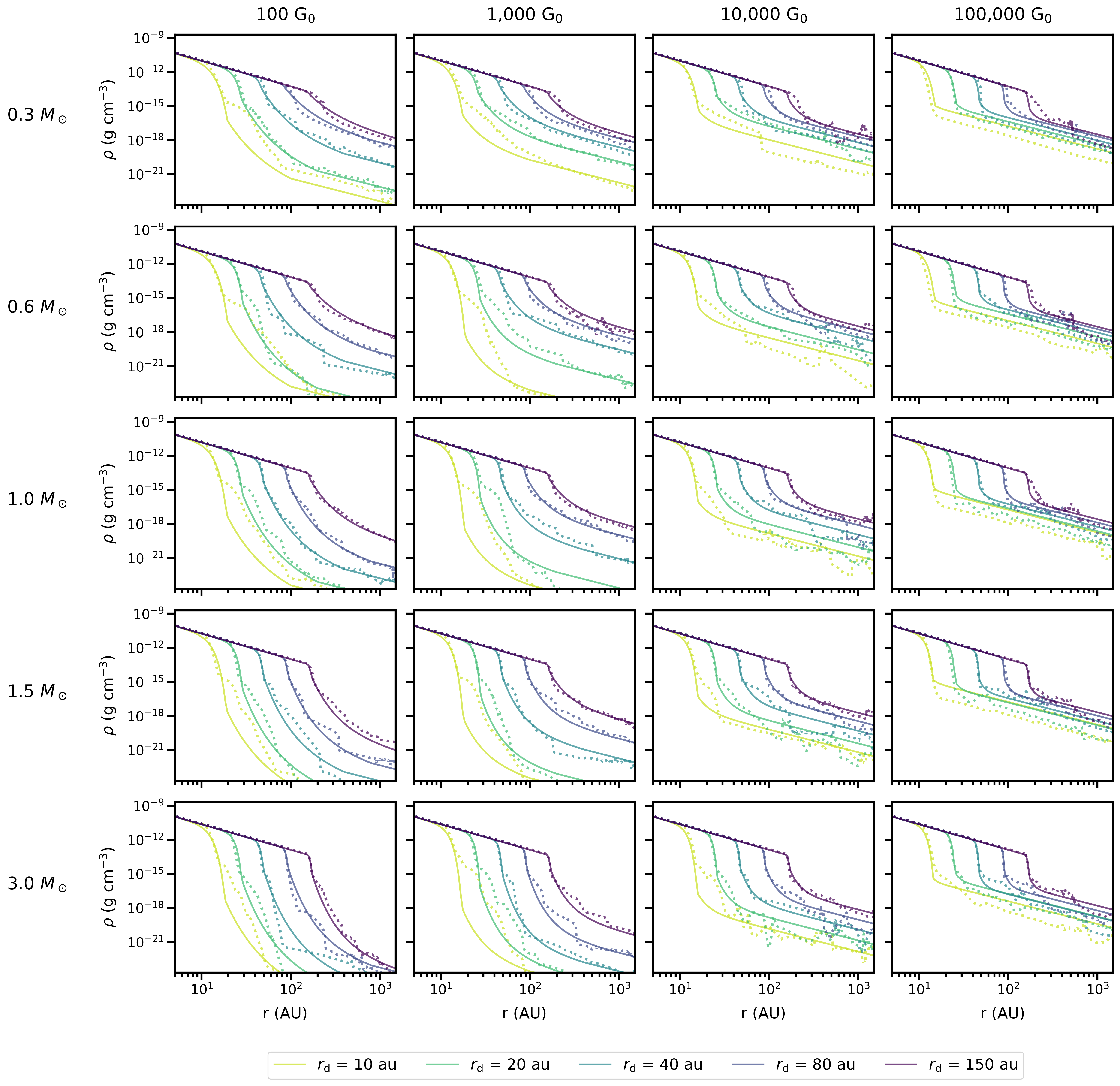}
\caption{Density profiles comparing our 1D parametric model (solid lines) with \textsc{fried} models (dotted lines) across different stellar masses and external FUV field strengths. Columns show increasing FUV field strength (100, 1000, 10,000, and 100,000 G$_0$). Rows represent different stellar masses (0.3, 0.6, 1.0, 1.5, 3.0 $M_\odot$). Each subplot displays four different disk sizes ($r_\text{d} = 20, 40, 80, 150$ au) represented by different colours. All models use a fixed surface density scaling parameter of $\Sigma_\text{1au} = 1000$ g cm$^{-3}$.}
\label{fig_1d_grid}
\end{figure*}

\subsubsection{Parameter calibration and validation}
\label{subsec:validating_1D}

To calibrate the free parameters of our parametric model, we compare directly against the \textsc{fried} grid of hydrodynamical simulations \citep{haworth_2018_fried, haworth_2023_fried2}. The \textsc{fried} models solve the full PDR–hydrodynamic problem using the \textsc{torus-3dpdr} code \citep{bisbas_2015, harries_2019}, computing self-consistent flow structures and mass-loss rates by imposing a zero-velocity boundary condition at the disk outer edge.

For the calibration, we selected a representative subset of the \textsc{fried} grid spanning stellar masses from $0.3$–$3.0\;M_\odot$, disk radii from 10–150 au, surface densities from $10^0$–$10^4\ \text{g cm}^{-2}$, and external FUV fields from $10^2$–$10^5$ G$_0$. These combinations produce mass-loss rates ranging from $\sim10^{-17}$ to $10^{-5}\;M_\odot\;\text{yr}^{-1}$, and comprise a total of 600 hydrodynamic models.

We adopted a Markov Chain Monte Carlo (MCMC) approach using the \textsc{emcee} ensemble sampler \citep{emcee_foreman_mackey_2013} to determine the best-fitting values of the free parameters $A$, $\alpha$, $\beta$, $\varepsilon$, $p$, and $q$ introduced in the previous section. Because the hydrodynamic density profiles exhibit significant numerical noise, particularly in the low-density `plateau’ region, we fit the parameters that control the exponential taper and those controlling the plateau shape in two separate stages.

In the first stage, we fit only the parameters governing the exponential taper, comparing only density values in regions with $r > r_\text{d}$ and $\rho > 10^{-16}\,\text{g cm}^{-3}$. This excludes the low-density `plateau' region, allowing the taper parameters to be constrained cleanly. At each MCMC iteration, a trial parameter set $\theta$ was drawn from uniform priors spanning $[-2, +2]$ for each parameter. For this trial $\theta$, we computed the parametric density profile corresponding to the physical inputs of every model in the selected subset of the \textsc{fried} grid. The log-likelihood contribution from each model was then evaluated and summed. We define the log-likelihood as
\begin{equation}
    \ln \mathcal{L}(\theta) = -\frac{1}{2} \sum_j \sum_i \bigg[\frac{\ln \rho_{\text{\tiny PARAMETRIC}} (r_i|\theta,j) - \ln \rho_{\text{\tiny FRIED}}(r_i)}{\sigma}\bigg]^2
\end{equation}
where the first sum runs over all \textsc{fried} models and the second sum runs over the radial points within each model. The term $\sigma$ represents the assumed uncertainty in the logarithmic densities. We adopt a constant fractional error by setting $\sigma=\ln(1+f)$ with $f=1$, which corresponds to allowing the parametric and hydrodynamical densities to differ by up to a factor of two. This level of uncertainty is appropriate given the numerical noise and sharp gradients present in the hydrodynamical profiles.

We ran the MCMC for 20,000 iterations (excluding 1000 burn-in steps) using 20 walkers, which yielded well-converged posterior distributions with best-fitting values $A=0.77$, $\alpha=0.10$, $\beta=0.78$, $\varepsilon=0.07$ (Figure \ref{fig_corner_plot}). The posterior distributions reveal a partial degeneracy between $A$ and $\beta$. This reflects the empirical behaviour of the \textsc{fried} models, in which the taper steepness varies most strongly with disk radius, while its dependence on stellar mass and external FUV field strength is comparatively weaker. As a result, changes in the baseline normalisation $A$ can be partially compensated by changes in the radial scaling without significantly altering the taper behaviour over the disk radii used to constrain the fit.

In the second stage, these four taper parameters were fixed at their best-fit values, and we fitted only the remaining parameters $p$ and $q$ that control the plateau region of the density profile. In this stage the full density profile was used when evaluating the likelihood. The resulting best-fit values were $p=0.2$ and $q=0.4$.

Figure \ref{fig_1d_grid} demonstrates the quality of the agreement between our parametric model (using these fitted parameters) and the \textsc{fried} simulations across the full parameter space. The parametric model accurately reproduces the density structure across a diverse range of stellar masses, disk sizes and masses, and external FUV field strengths.

The quality of agreement is weaker for very truncated (10\,au) disks. This is because the \textsc{fried} calculations struggle to obtain true steady state solutions in that regime, where the disk outer edge is comparatively bound and there are steep density gradients between the disk boundary condition and wind. Material also moves into a form of hydrostatic equilibrium near the disk outer edge which gives some of the build up seen in Figure \ref{fig_1d_grid}  in those cases. Within \textsc{fried} that was justified because the main quantity of interest is the mass loss rate, which in those cases is always low ($<10^{-10}\,$M$_\odot$\,yr$^{-1}$) and was estimated using a time averaged value of the pseudo-steady flow that results. Our physically motivated parametrised wind structure provides a cleaner, sensible density profile in this (weak wind) regime with a mass loss rate consistent with \textsc{fried}, without artefacts due to the numerical limitations of \textsc{fried}.

\subsection{2D model extension}

\subsubsection{Model construction}

While our 1D model effectively reproduces hydrodynamical simulations across a wide range of parameters, a full 2D treatment is necessary to capture the spatial structure required for detailed chemical modelling and comparison with observations. The extension to two dimensions introduces several additional complexities. Most notably, the disk vertical structure becomes significantly more extended under strong external FUV irradiation compared to standard isothermal prescriptions. Furthermore, the transition from disk to wind must now be handled across the entire two-dimensional disk surface rather than at a single radial location.

Our 2D model builds systematically upon the 1D framework while addressing these additional challenges. The model construction can be divided into three stages: (i) we establish the disk density structure under hydrostatic equilibrium, accounting for external FUV heating; (ii) we implement a spherically diverging wind emanating from a focal point within the disk; and (iii) we bridge these components with a transition region that captures the gradual density decline from the disk surface into the wind. As in the 1D case, the final density at any location is determined by taking the maximum of these three components.

We begin by constructing the disk density structure using the same surface density and temperature prescriptions as the 1D model. The midplane density is then given by:
\begin{equation}
    \rho_0 (r)= \frac{1}{\sqrt{2\pi}} \frac{\Sigma(r)}{h(r)}
\end{equation}
and the vertical structure initially assumes a Gaussian profile:
\begin{equation}
    \rho_\text{disk} = \rho_0 \exp\left(-z^2/2h^2\right)
\end{equation}
where $z$ represents the vertical height above the midplane.

This isothermal approximation breaks down under strong external irradiation, which can heat the disk surface to temperatures far exceeding the midplane value. To obtain a more realistic vertical structure, we solve for hydrostatic equilibrium by determining both the FUV radiation field and the resulting temperature structure. We achieve this through an iterative procedure that cycles through the following steps until convergence:

\emph{1. FUV field and optical depth.} We compute the total FUV field at each location by combining contributions from the central star and the external radiation field. The stellar FUV luminosity is estimated using empirical scaling relations \citep{eker_2018} (see Appendix \ref{appendix:fuv_fraction}), while the external field is applied isotropically. We calculate the FUV optical depth using a representative opacity $\kappa_\text{FUV} = \sigma_\text{FUV}/(\mu m_\text{p})$, where $\sigma_\text{FUV}$ is the FUV absorption cross-section and $\mu$ is the mean molecular weight. Both parameters vary by region: in the dust-depleted wind we adopt $\sigma_\text{FUV}=2.7 \times 10^{-23}$ cm$^2$ and $\mu=1.3$, while in the disk we adopt $\sigma_\text{FUV}=8 \times 10^{-22}$ cm$^2$ and $\mu=2.3$ (following e.g. \citealt{facchini_2016, haworth_clarke_2019, paine_2025}). We evaluate the optical depth along three distinct paths: vertically downward from the upper boundary, radially inward from the outer boundary, and radially outward from the central star. The optical depth at each location is then taken as the minimum of these three values.

\begin{figure}
\centering
\includegraphics[clip=,width=1\linewidth]{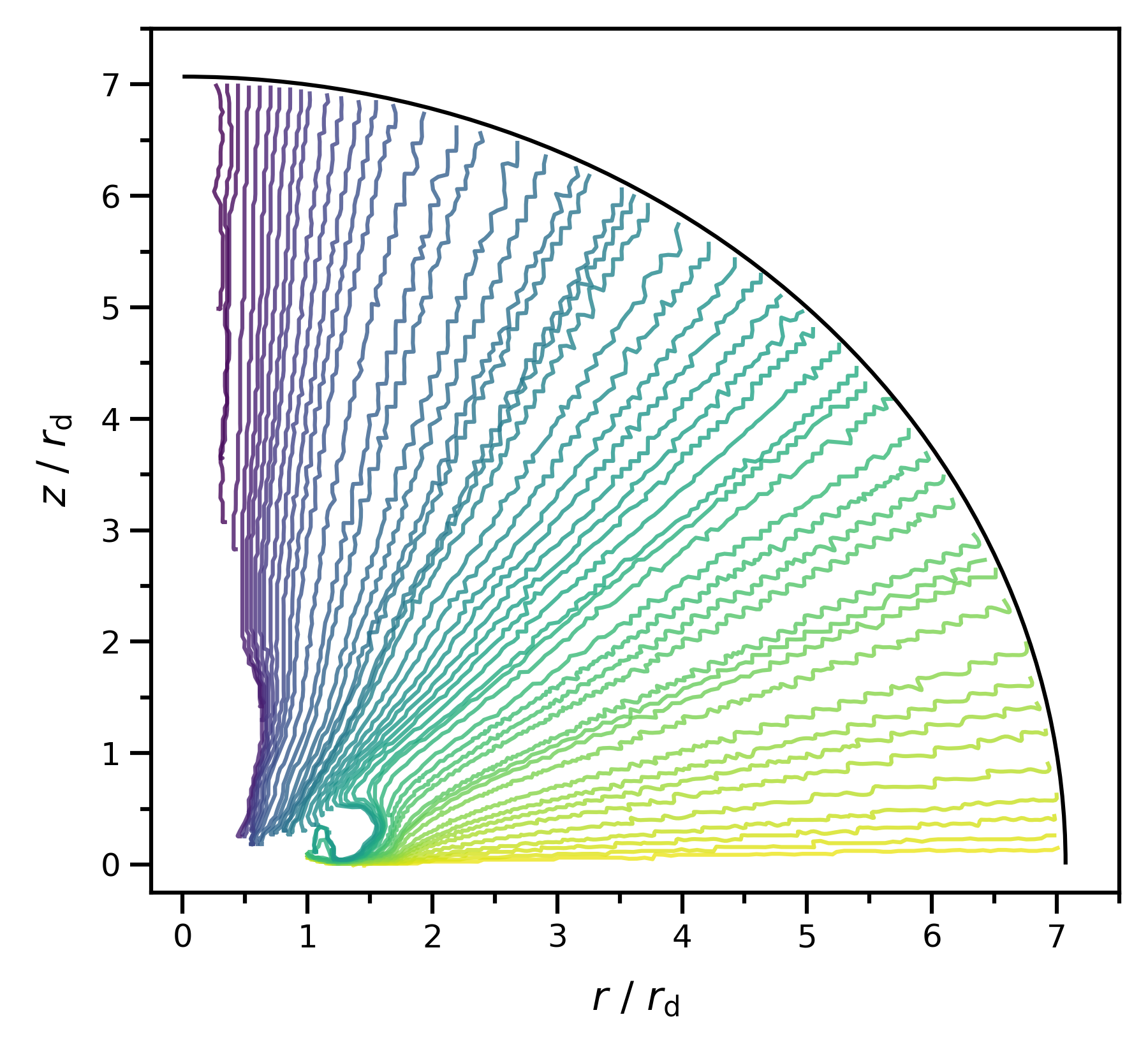}
\caption{Streamlines from a representative photoevaporation model with disk radius $r_\text{d} = 100$ au and external FUV field $F_\text{FUV} = 5000 \; G_0$. Streamlines are traced backward from a spherical surface in the outer wind region to their origins at the disk surface. The convergence of these streamlines identifies a focal point on the midplane located at approximately $r_\text{focal}=0.5 r_\text{d}$.}
\label{fig_streamlines}
\end{figure}

\emph{2. Temperature structure.} Using the calculated optical depth, we define the disk surface as the $\tau_\text{FUV}=1$ contour. Above this surface (where $\tau_\text{FUV}<1$), the gas is directly exposed to FUV radiation and we set the temperature to $T_\text{PDR}$ (Equation \ref{eq:t_pdr}). Below the surface, the temperature transitions smoothly from $T_\text{PDR}$ to the midplane temperature $T_\text{mid}$:
\begin{equation}
    T = T_\text{mid} + (T_\text{PDR} - T_\text{mid}) \times \beta
\end{equation}
where the transition function $\beta$ varies from 0 at the midplane to 1 at the surface:
\begin{equation}
    \beta = \frac{1}{2} \left[1 + \tanh\left(k \left(\frac{z}{z_\text{surface}} - \frac{1}{2}\right)\right)\right]
\end{equation}
with $k$ controlling the steepness of the vertical temperature gradient (following a similar approach to e.g. \citealt{williams_best_2014}).

\begin{figure*}
\centering
\includegraphics[clip=,width=1\linewidth]{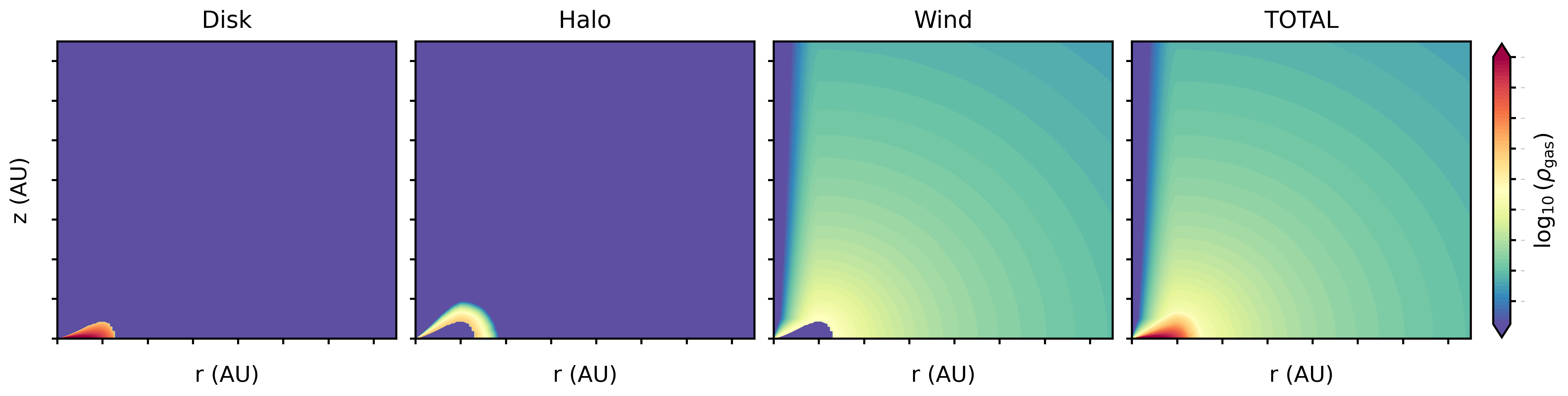}
\caption{Schematic illustration of the three main components that constitute our 2D parametric model. \emph{First panel:} The disk component, following hydrostatic equilibrium with a vertical density profile modified by external FUV heating. \emph{Second panel:} The disk-wind transition region, or `halo', which provides a smooth bridge between the disk surface and the wind through an exponential taper extending outward from the disk surface. \emph{Third panel:} The spherically diverging photoevaporative wind, emanating from a focal point located at approximately $0.5\;r_{\rm d}$ on the midplane. \emph{Fourth panel:} The final density structure at any location in the full 2D model is determined by taking the maximum density among these three components.}
\label{fig_model_components}
\end{figure*}

\emph{3. Hydrostatic equilibrium.} With the temperature structure established, we solve for hydrostatic equilibrium in the vertical direction. In the thin-disk approximation ($z \ll r$), the vertical density gradient is governed by:
\begin{equation}
    \frac{d \ln \rho}{dz} = -\frac{\Omega^2 z}{c_s^2} - \frac{1}{T} \frac{dT}{dz}
\end{equation}
where $\Omega = \sqrt{GM_*/r^3}$ is the Keplerian angular velocity. At each radius, we integrate this equation upward from the midplane density to obtain the vertical density profile. We then renormalize the profile to match the prescribed surface density $\Sigma(r)$, ensuring mass conservation. With the updated density structure, we recalculate the optical depth and temperature, then repeat this iterative cycle until the density converges (typically within $\lesssim 20$ iterations).

Having established the disk density, we implement the spherically diverging photoevaporative wind. Analysis of 2D hydrodynamical simulations reveals that streamlines in the outer wind region, when traced back toward the disk, converge toward a focal point located at approximately $r_\text{focal} = 0.5 r_\text{d}$ on the midplane \citep{haworth_clarke_2019}. This behaviour is illustrated in Figure \ref{fig_streamlines}, which shows streamlines from a representative hydrodynamical model. We therefore construct a wind that diverges spherically from this focal point, with density:
\begin{equation}
   \rho_\text{wind}(d) = \frac{\dot{M}}{4\pi d^2 v_\text{gas}} 
\end{equation}
where $d$ is the distance from the focal point and $v_\text{gas} = c_\text{s}$ (Equation \ref{eq:c_s_wind}). The mass-loss rate $\dot{M}$ can either be specified directly or interpolated from the \textsc{fried} grid. In the latter case, we apply a scaling factor of $2$ to account for the systematic underprediction of mass-loss rates by 1D models relative to 2D simulations with identical initial conditions \citep{haworth_clarke_2019}.

As in the 1D case, directly connecting the wind to the disk creates an unphysical density discontinuity. In 2D hydrodynamical simulations, external irradiation produces an extended disk surface resembling a puffed up `halo-like' structure that naturally bridges the dense disk interior and tenuous wind. We model this transition using the $\tau_\text{FUV}=1$ contour to define the disk surface. Normal to this surface and extending outward into the wind, we apply the transition solution from the 1D model. This creates a smooth density bridge where material gradually transitions from the bound disk to the expanding wind, following the geometry of the irradiated disk surface while maintaining physical continuity with both components.

A final refinement addresses an additional feature observed in 2D hydrodynamical simulations: at small radii ($r \lesssim r_\text{d}$), the wind density becomes suppressed relative to the purely spherical divergence prescription. This occurs because material close to the midplane at small radii remains more strongly bound and is not part of any externally-driven outflow. We implement this suppression by applying a smooth tapering function that reduces the wind density in the inner regions, with the suppression strongest near the midplane and extending further in radius at higher altitudes. Specifically, the tapering region extends to $r_\text{edge}(z) = r_\text{d}$ for $z \geq r_\text{focal}$ and to $r_\text{edge} = r_\text{d} \cdot (z/r_\text{focal})$ for $z < r_\text{focal}$. Within this region, we apply a standard cosine tapering function that smoothly reduces the wind density from its full value at $r = r_\text{edge}$ to zero at $r=0$:
\begin{equation}
   f_\text{taper}(r,z) = 1 - \frac{\cos[\pi({r}/{r_\text{edge}})^2]}{2}
\end{equation}
The modified wind density is then $\rho_\text{wind} \rightarrow f_\text{taper}(r,z) \times \rho_\text{wind}$.

The final density at any point in the 2D model is determined by taking the maximum of all three components:
\begin{equation}
    \rho(r,z) = \max[\rho_\text{disk}(r,z), \rho_\text{halo}(r,z), \rho_\text{wind}(r,z)]
\end{equation}
The full model is illustrated schematically in Figure \ref{fig_model_components}, which shows each of the individual model components and their combined structure.

\subsubsection{Comparison with hydrodynamic simulations}
\label{subsubsec:validating_2d_model}

Validating our 2D parametric model against a large grid of hydrodynamical simulations is not feasible due to the substantial computational cost of such simulations. They require iterations between 2D PDR calculations with 3D line cooling and hydrodynamics until steady state (see below) and hence typically require weeks to months to complete. Indeed, this computational expense is precisely the motivation for developing this efficient parametric approach. We therefore validate our 2D model by first comparing it against a representative suite of four hydrodynamical simulations originally presented by \citet{ballabio_2023} and subsequently employed in chemical modelling studies by \citet{keyte_haworth_2025}.

These simulations were generated using the photochemical hydrodynamics code \textsc{torus-3d-pdr} \citep{bisbas_2015}, following methodology similar to \citet{haworth_clarke_2019}. \textsc{torus-3d-pdr} couples the radiation hydrodynamics scheme from \textsc{torus} \citep{haworth_harries_2012} with multidimensional radiative transfer, equilibrium chemistry, and thermal balance calculations from \textsc{3d-pdr} \citep{bisbas_2012}. We refer readers to those studies for detailed descriptions of the modelling procedures. The four simulations adopt identical input parameters except for the external FUV field strength, which takes values of 100, 500, 1000, and 5000 G$_0$, while all other parameters remain fixed at $M_*=1.0\;M_\odot$, $r_\text{d}=100$ au, and $\Sigma_\text{1au}=1000$ g cm$^{-2}$.

We initialize four corresponding parametric models using these same input parameters, with mass-loss rates extracted directly from the \textsc{torus-3d-pdr} simulations. By default, our 2D models adopt the $\gamma$ parametrization established through the 1D validation, where $\gamma$ controls both the shape and radial extent of the transition region between disk and wind. However, users may prefer to treat $\gamma$ as a free parameter for either manual adjustment or systematic optimization within larger Bayesian model fitting frameworks.

\begin{figure*}
\centering
\includegraphics[clip=,width=1\linewidth]{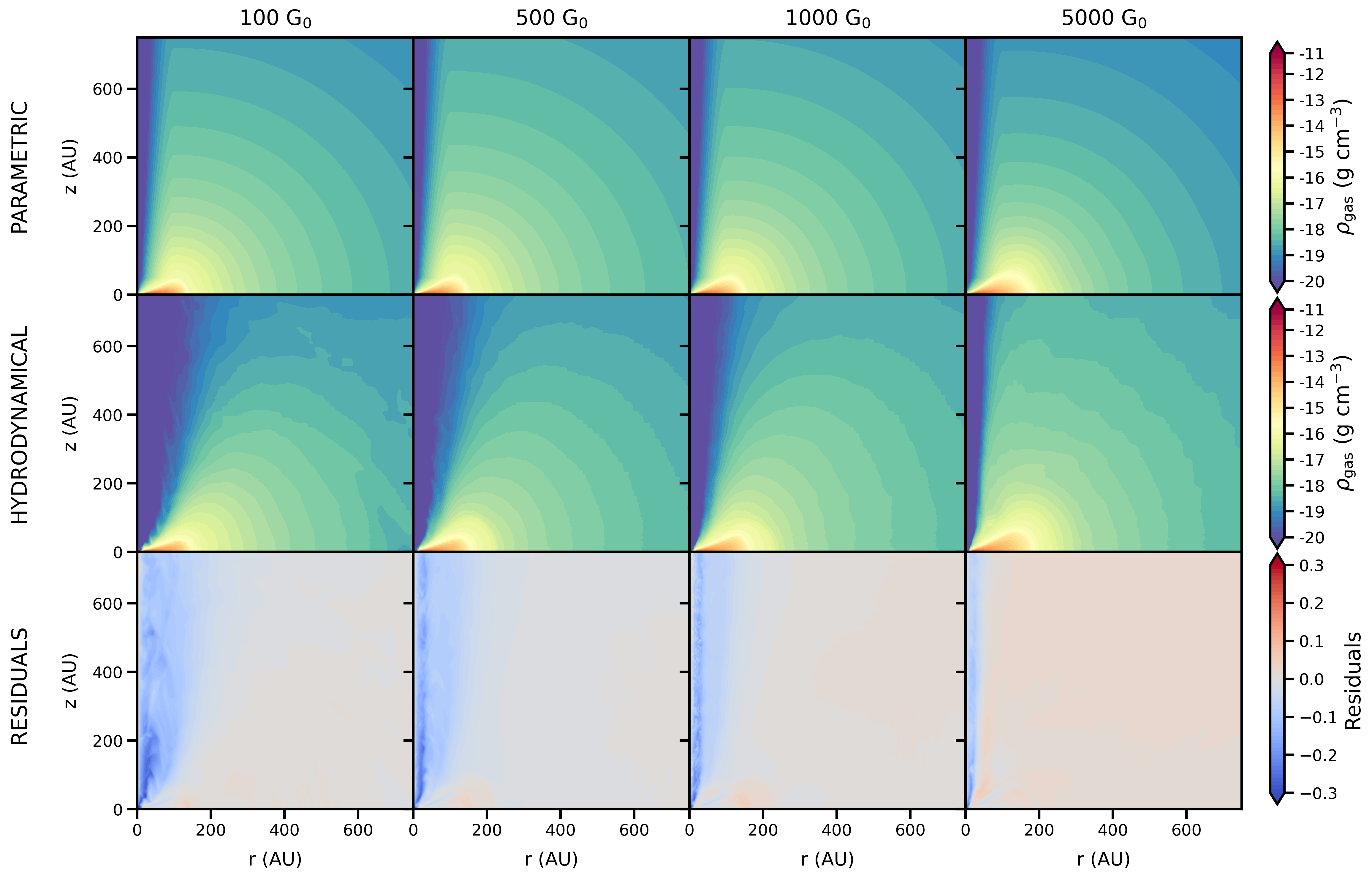}
\caption{Comparison between density structures from the parametric 2D model(top row) and hydrodynamic simulations (middle row) for four external FUV field strengths: 100, 500, 1000, and 5000 G$_0$. The bottom row shows the normalized logarithmic residuals (Equation \ref{eq:residuals}).}
\label{fig_2d_comparison}
\end{figure*}

Figure \ref{fig_2d_comparison} presents the comparison between our parametric models and the four hydrodynamical simulations, adopting $\gamma$ values of 4.75, 3.75, 3.5, and 2.25 for the 100, 500, 1000, and 5000 G$_0$ cases respectively. The figure displays parametric model densities in the top row, corresponding \textsc{torus} density structures in the middle row, and residuals in the bottom row. We quantify the agreement using logarithmic residuals weighted by local density:
\begin{equation}
   \delta = \frac{\log(\rho_{\text{\tiny HYDRO}}) - \log(\rho_{\text{\tiny PARAMETRIC}})}{|\max(\log(\rho_{\text{\tiny HYDRO}}), \log(\rho_{\text{\tiny PARAMETRIC}}))|}
   \label{eq:residuals}
\end{equation}
where the denominator acts as a weighting factor, emphasising discrepancies between the models in regions of higher density.

These results demonstrate that our parametric framework successfully reproduces the key morphological features of the hydrodynamical simulations across all four external FUV field strengths. The parametric models accurately capture both the spherically diverging wind structure in the outer regions and the smooth transition between wind and disk. The normalized residuals reveal that discrepancies are generally confined to $|\delta| \lesssim 0.1$ throughout most of the model domain, with parametric and hydrodynamical densities typically agreeing to within a factor of $<2$. The largest deviations occur in localized regions near the disk-wind interface, where residuals occasionally reach $|\delta| \sim 0.2$, corresponding to density differences of approximately a factor of $\sim 4$. These discrepancies arise primarily from our simplified treatment of the transition region, rather than solving the full hydrodynamic equations. Importantly, the quality of agreement remains consistent across the full range of external FUV field strengths from 100 to 5000 G$_0$, indicating that our parametric prescriptions successfully capture the physical scaling of disk-wind structure with external irradiation.

We emphasize that while these four 2D hydrodynamical models provide validation for the new two-dimensional geometric features such as the vertical structure and 2D wind implementation, the underlying physical prescriptions for disk properties and mass-loss rates have already been extensively validated against hundreds of 1D \textsc{fried} models (Section~\ref{subsec:validating_1D}). The 2D comparison therefore tests our extension of these well-calibrated parametrisations into two dimensions rather than the fundamental physics itself.

To assess model performance more broadly across the 2D parameter space, we generated density structures spanning the ranges listed in Table~\ref{table:model_grid_2d_validation}. This grid encompasses diverse disk masses, sizes, stellar masses, and external FUV field strengths, yielding 320 unique models. We evaluated each through visual inspection to identify conditions under which the parametric approach produces physically reasonable structures. Representative examples of model density structures across the full range of input parameters are presented in Appendix~\ref{appendix:examples_2d_parametric_model}.

The parametric model generates plausible disk/wind structures in the overwhelming majority of cases. However, we identified three specific issues that arise in a small subset of models, each with well-defined physical origins. First, difficulties emerge when disks become sufficiently optically thin that FUV radiation penetrates to radii interior to the nominal disk radius $r_\text{d}$. In such cases, defining the disk surface via the $\tau_\text{FUV}=1$ contour becomes problematic, and the physical meaning of the disk radius boundary condition itself becomes questionable. This can be avoided by enforcing a minimum surface density threshold of $\Sigma_\text{1au}\gtrsim10$ g cm$^{-2}$, ensuring the disk remains optically thick to FUV radiation within its nominal extent. This threshold still permits modelling of relatively low-mass disks; for example, a disk with $r_\text{d}=20$ au and $\Sigma_\text{1au} = 10$ g cm$^{-2}$ corresponds to $M_\text{disk} \approx 10^{-4}\;M_\odot$.

Second, in approximately 2\% of the 320 cases summarised in Table \ref{table:model_grid_2d_validation}, the halo-like transition region grows too massive, containing a substantial fraction of the total disk mass and producing unrealistically extended structures. This stems from our simplified disk-wind transition treatment, which was calibrated against 1D models. Extension to 2D presents challenges given the limited number of 2D simulations available for calibration. We mitigate this by rejecting models where $M_\text{halo} > 0.1 \; M_\text{disk}$, ensuring the transition region remains a minor component of the total structure.

Third, complications arise when a user-specified or \textsc{fried}-interpolated mass-loss rate produces wind structure with density at the wind base exceeding that at the disk surface, creating an unphysical discontinuity. In these cases, the wind density typically exceeds the disk density by factors of only approximately 2 to 3, and we scale the wind density downward accordingly. This adjustment is justified as such factors lie within typical mass-loss rate uncertainties.

These issues represent edge cases in the parameter space and do not compromise the model's utility for the vast majority of physically relevant disk configurations. Within the parameter space covered by our validation grid (Table~\ref{table:model_grid_2d_validation}), spanning stellar masses from 0.3 to 3.0~$M_\odot$, disk radii from 20 to 150~au, and external FUV fields from $10^2$ to $10^5$~G$_0$, the model quickly produces plausible and physically motivated 2D disk-wind density structures when the mitigation criteria described above are applied. We emphasise that our goal is not to perfectly reproduce hydrodynamical simulations, which remains challenging given the limited number of 2D models available for calibration, but rather to provide a computationally efficient framework for including realistic disk-wind representations across a wide parameter space. This enables chemical modelling and synthetic observations that would be impractical with full hydrodynamical calculations. We caution against extrapolating beyond these validated ranges without further testing, and note that users requiring high-fidelity reproduction of specific wind structures may need to treat $\gamma$ as a tunable parameter or validate against hydrodynamical models for their specific conditions of interest.

\begin{table}
\footnotesize
\caption{Parameter grid used to validate the 2D parametric model. We evaluate 320 models spanning all combinations of stellar mass, surface density, disk radius, and external FUV field strength. For each combination, the corresponding mass-loss rate is interpolated from the \textsc{fried} grid.}        
\label{table:model_grid_2d_validation}      
\centering
\setlength{\tabcolsep}{5pt} 

\begin{tabular}{l l l}     %
\hline\hline       
Parameter & Description & Value \\ 
\hline                    
   $M_*$                & Stellar mass                    & [0.3, 0.6, 1.0, 3.0] M$_\odot$     \\
   $\Sigma_\text{1au}$  & Surface density at $r$=$1$ au   & [10$^0$, 10$^1$, 10$^2$, 10$^3$, 10$^4$] g cm$^{-2}$    \\
   $R_\text{d}$         & Disk outer edge                 & [20, 40, 80, 100, 150] au               \\
   $F_\text{FUV}$       & Integrated FUV field            & [10$^2$, 10$^3$, 10$^4$, 10$^5$] G$_0$        \\
   $\dot{M}$            & Mass-loss rate                  & [$\sim 10^{-17}$ to $\sim 10^{-5}$] $M_\odot$ yr$^{-1}$   \\
\hline                  
\end{tabular}
\end{table}

\subsection{Code implementation}

We provide both the 1D and 2D parametric models as the lightweight Python package \textsc{puffin} ('Python Utility For Fuv Irradiated disk deNsities'). The package requires only standard scientific Python libraries (\texttt{NumPy}, \texttt{SciPy}, \texttt{Matplotlib}) and can generate full 2D density structures in seconds to minutes, compared to weeks or months for equivalent hydrodynamical simulations.

\textsc{puffin} supports flexible model construction with either user-specified mass-loss rates or automatic interpolation from the \textsc{fried} grid. Output density structures and coordinate grids are provided as \texttt{NumPy} arrays and can easily be exported to other formats compatible with widely-used chemical modelling codes. The package includes validation utilities that check for the edge-case issues described in Section \ref{subsubsec:validating_2d_model} and warn when input parameters fall outside the validated range.

The package is available via the Python Package Index (PyPI) and can be installed using the standard \texttt{pip} command:

\begin{verbatim}
pip install puffin_disk
\end{verbatim}

\noindent Comprehensive documentation, tutorials, and example notebooks are 
available at:
\begin{verbatim}
puffin.readthedocs.io
\end{verbatim}

\section{Applications to Disk Chemistry}
\label{sec:chemical_modelling}

Having established and validated our parametric framework, we now demonstrate its application to disk chemistry. The computational efficiency of our model enables systematic exploration across parameter space that would be prohibitively expensive with hydrodynamical simulations. Potential applications include developing observational diagnostics for spatially unresolved or marginally resolved systems, investigating how external irradiation affects volatile reservoirs and snowline locations relevant for planet formation, exploring chemical inheritance versus reset across evolutionary stages, and identifying molecular tracers that provide robust diagnostics of irradiation history. Such studies could address fundamental questions about how environment shapes the chemical evolution of planet-forming disks and the compositional diversity of resulting planetary systems.

As a first illustration of these capabilities, we investigate how variations in external FUV field strength influence the spatial distribution and abundance of CO in protoplanetary disks. CO serves as an ideal test case, since it is a fundamental chemical tracer, and has been widely detected in externally irradiated environments such as the Orion Nebula Cluster \citep{boyden_eisner_2020, boyden_eisner_2023}. This analysis demonstrates the practical utility of our model while providing insight into how the chemistry of disk-wind systems respond to varying external radiation fields.

\subsection{Chemical model setup}

We construct a grid of chemical models spanning the parameter space defined in Table~\ref{table:model_grid_2d_validation}. For each combination of stellar mass, disk mass, disk radius, and external FUV field strength, we first generate a 2D density structure using our parametric model, then use this structure as input to detailed thermochemical calculations.

\subsubsection{Physical structure}

The thermochemical modelling uses the \textsc{dali} code \citep{Bruderer2012, Bruderer2013}, a well-established 1+1D computational framework that has been widely applied to model chemistry in protoplanetary disks \citep[e.g.][]{kama_2016b, van_der_marel_2016, Facchini17, fedele_2017, cazzoletti_2018, maps_7_bosman_2021, long_2021, leemker_2022, keyte_2023, zagaria_2023, stapper_2024, vlasblom_2024, paneque_carreno_2025, vaikundaraman_2025}. \textsc{dali} requires as input a parametrised gas and dust density distribution along with a stellar spectrum. The code then employs Monte Carlo radiative transfer to compute the UV radiation field and dust temperature throughout the disk. UV photons originate from two sources: the central star (characterized by the input stellar spectrum) and the interstellar radiation field (ISRF), which is implemented by propagating photons uniformly inward from a virtual sphere encompassing the computational domain. The resulting dust temperature provides an initial estimate for the gas temperature, starting an iterative procedure in which the chemistry is evolved time-dependently.

Our parametric model provides only the gas density structure, deliberately omitting dust prescriptions to maintain flexibility and simplicity. This design choice allows users to implement whichever dust treatment best suits their specific application. Here, we adopt the approach of \citet{haworth_clarke_2019}, in which photoevaporative winds exhibit both a reduced maximum grain size ($a_\text{max}$) and a lower dust-to-gas mass ratio ($\Delta_\text{d/g}$) compared to their parent disk, as expected theoretically \citep{facchini_2016, owen_altaf_2021, paine_2025}.

We set $\Delta_\text{d/g} = 0.01$ in the disk and $\Delta_\text{d/g} = 3 \times 10^{-4}$ in the wind. The wind dust consists of a single population of small grains ($0.005$--$1\,\mu$m) following the standard ISM grain size distribution \citep{mathis_1977}. The disk region contains two populations: small grains identical to those in the wind, and large grains ($0.005$--$1000\,\mu$m) following the same size distribution. The fraction of the total dust mass within the large grain population is given by the parameter $f$.

Dust settling is implemented following the approach of \citet{dalessio_2006}, which treats small and large grain populations with different vertical scale heights. The small grains (0.005--1~$\mu$m) share the same scale height as the gas, while large grains (0.005--1000~$\mu$m) have a scale height reduced by a factor $\chi < 1$.  The vertical density profiles for each population are then given by:
\begin{equation}
    \rho_\text{dust,small} = \frac{(1-f)\Sigma_\text{dust}}{\sqrt{2\pi}rh} \exp\left[-\frac{1}{2}\left(\frac{\pi/2-\theta}{h}\right)^2\right]
\end{equation}
\begin{equation}
    \rho_\text{dust,large} = \frac{f\Sigma_\text{dust}}{\sqrt{2\pi}r\chi h} \exp\left[-\frac{1}{2}\left(\frac{\pi/2-\theta}{\chi h}\right)^2\right]
\end{equation}
where $r$ is the radial distance and $\pi/2 - \theta \sim z/r$. Since our parametric model directly provides the gas density structure, we first compute $h(r)$ from the gas density at each radius, then apply the above prescriptions to construct the dust distribution. We adopt typical values of $f = 0.9$ and $\chi = 0.2$ \citep[e.g.][]{kama_2016a, fedele_2017, trapman_2020b, leemker_2022, keyte_2024b, keyte_2024a, stapper_2024}.

The computational grid comprises 200 logarithmically-spaced cells in both radial and vertical directions (40,000 cells total), extending from $r = 0$ to $r = r_\text{out} = 8 r_\text{d}$ in both dimensions. This resolution adequately captures the disk structure while encompassing the full spatial extent of the photoevaporative wind. Logarithmic scaling in the vertical direction is required to provide sufficient resolution at the midplane for accurately treating the vertical distribution of settled large grains.

An important caveat concerns the auxiliary quantities computed during construction of the parametric density structure. While that model includes simplified prescriptions for gas temperature, UV field strength, and optical depth, these are used solely to establish the density distribution and should not be employed directly in chemical calculations. The density structures themselves are based on physically motivated parameterisations and have been shown to be consistent with hundreds of 1D and our handful of available 2D simulations (Sections~\ref{subsec:validating_1D} and \ref{subsubsec:validating_2d_model}). They have also been shown to result in plausible disk/wind structures over a wide parameter space (Appendix \ref{appendix:examples_2d_parametric_model}). However, the associated temperature and radiation field prescriptions are intentionally simplified. \textsc{dali} instead computes these quantities using full radiative transfer and thermal balance.

\subsubsection{Stellar parameters}

Our model grid spans stellar masses of 0.3, 0.6, 1.0, and 3.0~$M_\odot$. For each stellar mass, we generate a blackbody spectrum with effective temperature and luminosity determined from empirical scaling relations \citep[][see Appendix~\ref{appendix:fuv_fraction}]{eker_2018}. The X-ray emission is modelled as a thermal spectrum with temperature $T_X = 7 \times 10^7$~K over the 1--100~keV energy range, adopting a representative X-ray luminosity of $L_X = 7.94 \times 10^{28}$~erg~s$^{-1}$ typical of solar-mass T Tauri stars. Although observational studies indicate that X-ray luminosity scales with stellar mass \citep{flaischlen_2021}, we deliberately hold $L_X$ constant for simplicity. This approach also allows us to isolate the influence of external FUV irradiation on the disk structure, and more easily disentangle the effects of external irradiation from variations in the stellar radiation field.

\subsubsection{Chemical network}

The chemical evolution is calculated using a network based on a subset of UMIST~06 \citep{woodall2007}, as presented in \citet{visser_2018}. This network comprises 134 species connected through 1855 reactions, encompassing H$_2$ formation on dust grains, freeze-out and thermal desorption, gas-phase reactions, photodissociation and photoionization, X-ray-induced processes, cosmic-ray-induced chemistry, PAH charge exchange and hydrogenation, and reactions involving vibrationally excited H$_2$. Non-thermal desorption is included for a limited set of volatile species (CO, CO$_2$, H$_2$O, CH$_4$, NH$_3$, N$_2$), while grain surface chemistry is restricted to hydrogenation of simple species (C, CH, CH$_2$, CH$_3$, N, NH, NH$_2$, O, OH, CN). A complete description of these processes is provided in \citet{Bruderer2012}. We adopt initial molecular abundances based on \citet{ballering_2021}, which represent a typical protoplanetary disk with Solar-like composition inherited from its parent molecular cloud (Table~\ref{table:initial_abundances}). The chemistry is evolved for 1~Myr, representative of typical ages for the systems under investigation.

\begin{table}
\caption{Initial abundances for all chemical models presented in Section \ref{sec:chemical_modelling}, adapted from \citet{ballering_2021}. These values represent a typical protoplanetary disk with Solar-like C/O=0.5 and composition inherited from the parent molecular cloud.}        
\label{table:initial_abundances}      
\centering
\begin{tabular}{l l}     
\hline\hline       
Species & Abundance (X/H)  \\ 
\hline                    
H$_2$    & $5.00 \times 10^{-1}$    \\
He       & $9.00 \times 10^{-2}$    \\
H$_2$O   & $1.00 \times 10^{-4}$    \\
CO       & $1.00 \times 10^{-4}$    \\
N$_2$    & $3.51 \times 10^{-5}$    \\
C$^+$    & $1.00 \times 10^{-5}$    \\
NH$_3$   & $4.80 \times 10^{-6}$    \\
HCN      & $2.00 \times 10^{-8}$    \\
H$_3^+$  & $1.00 \times 10^{-8}$    \\
S$^+$    & $9.00 \times 10^{-9}$    \\
HCO$^+$  & $9.00 \times 10^{-9}$    \\
C$_2$H   & $8.00 \times 10^{-9}$    \\
Mg$^+$   & $1.00 \times 10^{-11}$   \\
Si$^+$   & $1.00 \times 10^{-11}$   \\
Fe$^+$   & $1.00 \times 10^{-11}$   \\
\hline                  
\end{tabular}
\end{table}

\subsection{Fiducial model results}
\label{subsec:co_abundance_study}

Having constructed the full model grid, we begin our analysis by focusing on a fiducial set of four models that isolate the effects of external FUV irradiation while holding all other parameters constant. These models are intended to represent typical externally irradiated protoplanetary disks, similar to those found in regions such as the Orion Nebula Cluster. The fiducial parameters are a stellar mass of $M_* = 0.6~M_{\odot}$, a disk radius of $r_\text{d} = 40$ au, and a surface density normalization of $\Sigma_{1\rm{au}} = 1000$ g cm$^{-2}$ (corresponding to $M_\text{disk} \sim 3 \times 10^{-2} M_\odot$).

\begin{figure*}
\centering
\includegraphics[clip=,width=1\linewidth]{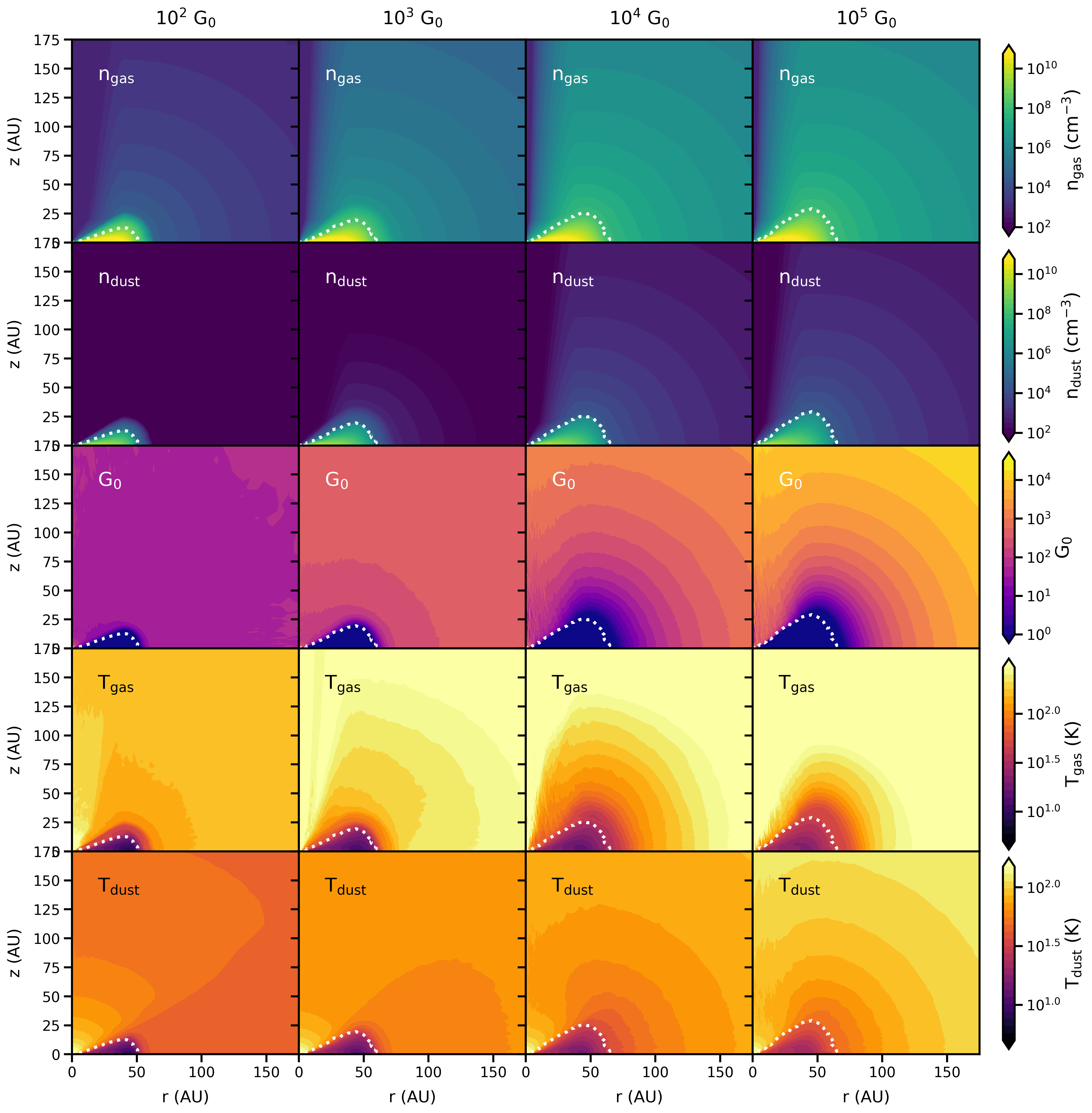}
\caption{Comparison of four models with varying external FUV field strengths ranging from $10^2$ to $10^5$ G$_0$ (columns). All other model parameters are fixed at $M_*=0.6\;M_\odot$, $\Sigma_\mathrm{1au}=1000$ g cm$^{-2}$, and $r_\mathrm{d}=40$ au. Rows display (from top to bottom): gas number density ($n_\mathrm{gas}$), dust number density ($n_\mathrm{dust}$), gas temperature (K), dust temperature (K), and local FUV field strength in Habing units (G$_0$). Dotted lines indicate the disk/wind transition boundary.}
\label{fig_fiducial_structure}
\end{figure*}

\subsubsection{Density and radiation fields}

Figure \ref{fig_fiducial_structure} presents the complete two-dimensional density/temperature structure for each model. The top two rows display the gas and dust number densities, with dashed contours marking the $\tau_{\rm FUV} = 1$ surface that delineates the disk-wind boundary. Although all models initialise with identical surface density profiles, the hydrostatic equilibrium calculation under external irradiation produces systematically more extended vertical structures as the external FUV field strength increases. At $100$ G$_0$, the disk maintains a relatively compact atmosphere, with the $\tau_{\rm FUV} = 1$ contour reaching $z/r \sim 0.3$ at the outer edge. The disk atmosphere becomes increasingly puffed up as the external FUV field increases, pushing the $\tau_{\rm FUV} = 1$ contour up to $z/r \sim 0.5$ when the external FUV field reaches $10^5$ G$_0$

The density structure of the photoevaporative wind exhibits corresponding enhancements. Mass-loss rates increase from $\sim 8 \times 10^{-10} M_\odot$ yr$^{-1}$ at $100$ G$_0$, to $\sim 3 \times 10^{-6} M_\odot$ yr$^{-1}$ at $10^5$ G$_0$, producing systematically higher wind densities.

The third row of Figure \ref{fig_fiducial_structure} displays the local FUV field strength in Habing units. In models with low external fields, stellar FUV emission dominates at the $\tau_{\rm FUV} = 1$ surface within the inner few au. At larger radii, the external radiation field becomes the primary FUV source above the disk surface. However, the enhanced photoevaporative wind density at high external FUV fields provides substantial attenuation, reducing to FUV flux by up to several orders of magnitude before it reaches the disk surface. Once FUV radiation penetrates the disk surface, it undergoes rapid attenuation through the increasingly dense gas and dust, falling well below interstellar levels within the disk interior for all models.

\begin{figure}
\centering
\includegraphics[clip=,width=1\linewidth]{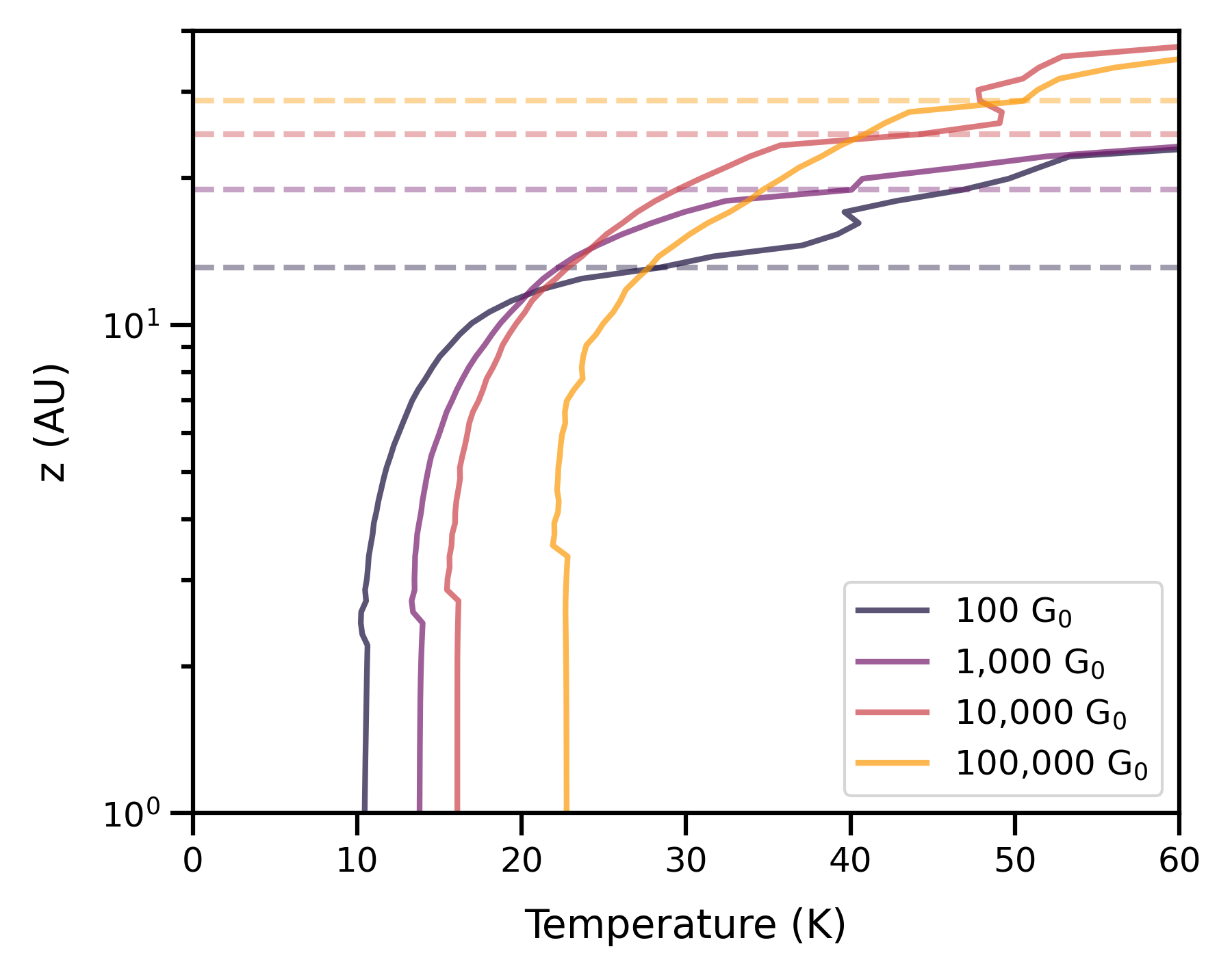}
\caption{Gas temperature vs. height for the four models with varying external FUV field strength (solid lines). Profiles were extracted from the outer edge of the disk at $r=r_\text{d}=40$ au. Dotted lines represent the approximate location of the disk-wind boundary. Stronger external FUV fields increase midplane temperatures through indirect heating, whereby FUV photons absorbed in the warm photoevaporative wind are re-radiated as infrared radiation, which penetrates deeper into the disk interior.}
\label{fig_vertical_temperature}
\end{figure}

\subsubsection{Thermal structure}
\label{subsec:model_thermal_structure}

The resulting thermal structure, shown in the bottom two rows of Figure \ref{fig_fiducial_structure}, exhibits several distinctive features. In the outer photoevaporative wind, gas temperatures reach several hundred Kelvin, primarily driven by three heating mechanisms: photoelectric heating from FUV photons absorbed by dust grains and PAHs, collisional de-excitation of FUV-pumped H$_2$, and H$_2$ photodissociation. These elevated temperatures decline sharply toward the nominal disk surface, where densities increase rapidly, falling to $\sim 50$ K at the $\tau_{\rm FUV} = 1$ surface.

Within the disk interior, a more subtle but important effect is evident. Gas temperatures increase systematically with external FUV field strength, despite heavy attenuation preventing FUV photons from directly penetrating to these depths. This is particularly clear at the disk outer edge, where midplane temperatures rise from $\sim 10$ K under a 100 G$_0$ field to $\sim 23$ K under a $10^5$ G$_0$ field, which is sufficient to exceed the canonical CO sublimation temperature of $\sim 18$ K (Figure \ref{fig_vertical_temperature}). This additional heating operates through an indirect mechanism, where FUV photons absorbed in the wind are re-radiated as infrared radiation, which can penetrate more deeply into the disk interior, warming both gas and dust \citep{keyte_haworth_2025}.

Dust temperatures (fifth row) follow similar trends but with some important differences. In the wind region, dust temperatures are systematically lower than gas temperatures, while in the disk interior the dust and gas remain well-coupled thermally. At the midplane, dust temperatures mirror those of the gas, increasing with stronger external FUV fields. These temperature variations have direct implications for CO gas- and ice-phase abundances, which we examine in the following section.

\begin{figure*}
\centering
\includegraphics[clip=,width=1\linewidth]{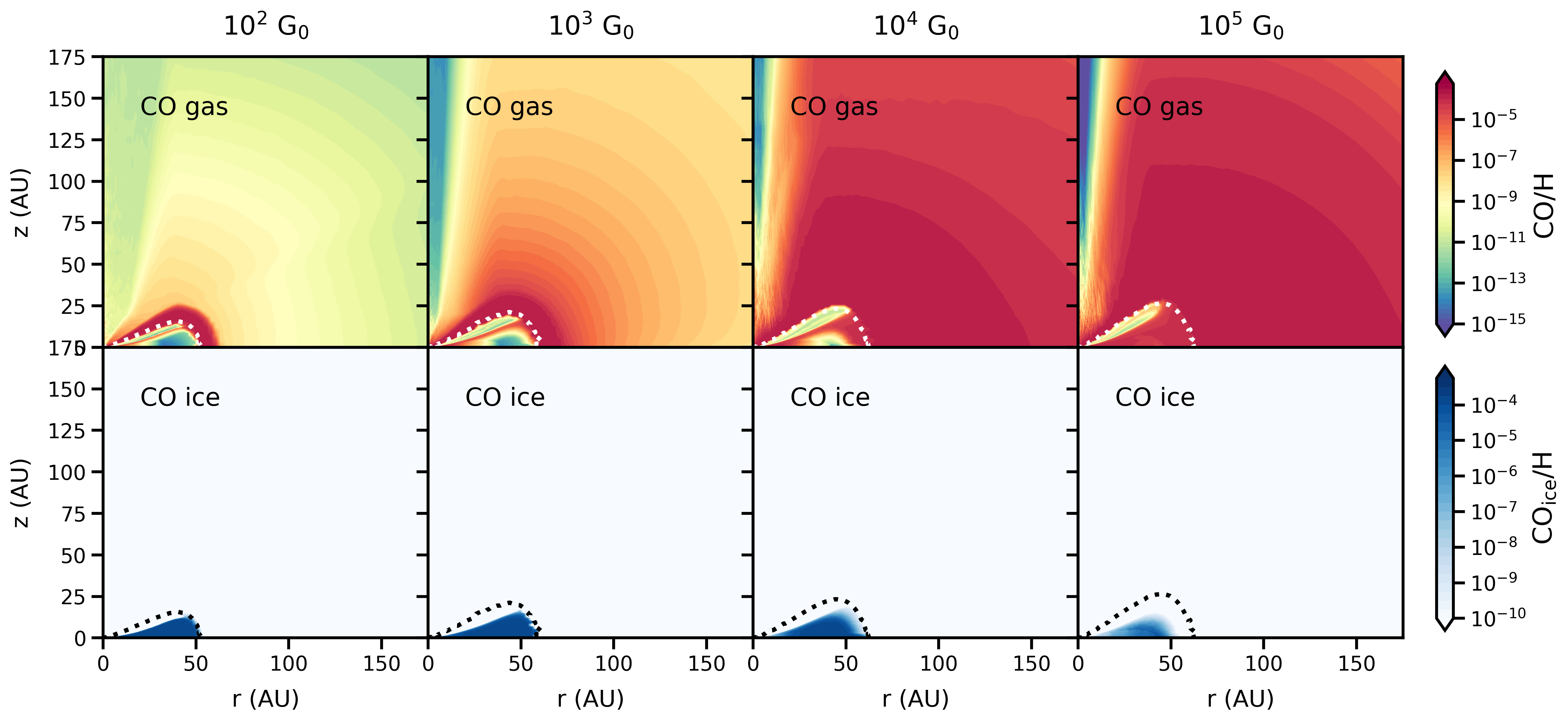}
\caption{CO abundances in the fiducial model ($M_* = 0.6\,M_\odot$, $\Sigma_{1\mathrm{au}} = 1000$ \,g\,cm$^{-2}$) as a function of external FUV field strength. \emph{Top row:} gas-phase CO abundance (CO/H). \emph{Bottom row:} ice-phase CO abundance (CO$_\mathrm{ice}$/H). Each column shows a different external field strength ($10^2$--$10^5$ G$_0$). Dotted lines trace the disk-wind boundary. With increasing external radiation, gas-phase CO abundance rises in the disk interior due to enhanced thermal desorption of CO ice, while the ice reservoir is correspondingly depleted. The photoevaporative wind layer (above the dotted line) also becomes progressively enriched in CO gas at higher FUV field strengths.}
\label{fig_fiducial_CO}
\end{figure*}

\subsubsection{CO gas and ice distribution}

Figure \ref{fig_fiducial_CO} illustrates the impact of increasing the external FUV field strength on the abundance of CO gas (top row) and ice (bottom row) for each model. At the lowest external FUV field strength ($100$ G$_0$), gas-phase CO is primarily abundant in the disk surface layers, which maintains temperatures above the CO sublimation temperature. Within the disk interior, lower dust temperatures lead to efficient freezeout, producing a large CO ice reservoir.

As external FUV irradiation increases to $10^3$ G$_0$, gas-phase CO becomes more prominent in the photoevaporative wind and slightly enhanced in the disk's upper layers, though a substantial ice reservoir still persists. At $10^4$ G$_0$, CO becomes highly abundant throughout the wind and survives in the gas phase deeper into the disk interior. The ice reservoir remains significant but begins eroding from the outside-in, reducing both its vertical and radial extent. 

Under the most intense irradiation of $10^5$ G$_0$, temperatures remain well above the CO sublimation point throughout most of the disk, reaching $\sim 30$ K at the disk outer edge. Gas-phase CO abundances approach CO/H$\sim 10^{-4}$ nearly everywhere (the maximum abundance, reflecting incorporation of the bulk carbon and oxygen budget), while the ice reservoir becomes severely depleted. The infrared heating mechanism described in Section \ref{subsec:model_thermal_structure} drives ice sublimation progressively inward: vertically downward from the disk surface and radially inward from both the inner and outer disk edges. This leaves the highest ice abundances concentrated near the midplane at the disk outer edge ($r_\text{d}$). Meanwhile, gas-phase CO also maintains an abundance of CO/H$\sim 10^{-4}$ throughout the photoevaporative wind.

A notable feature visible in Figure \ref{fig_fiducial_CO}, particularly at higher external FUV field strengths, is a narrow layer of reduced CO abundance situated just below the disk-wind boundary. This layer lies between two regions of high CO abundance (the warm wind above and the indirectly-heated midplane below), yet is itself strongly depleted. This comes about because the layer occupies a narrow range of physical conditions where CO destruction is uniquely efficient. First, X-ray ionisation rates are sufficiently high to drive ion–molecule chemistry. Second, dust temperatures are warm enough ($\sim$25–30 K) to keep CO in the gas phase. Third, and crucially, temperatures remain low enough that the organic products of CO destruction freeze out onto grain surfaces. Only within this intermediate transition zone do all three conditions occur simultaneously.

The destruction mechanism proceeds as follows. Gas-phase CO undergoes frequent reactions with ions such as H$_3^+$, but these primarily form HCO$^+$, which rapidly returns carbon to CO through reactions with PAHs and electrons. However, a fraction of CO reacts with He$^+$ (produced by stellar X-rays) via He$^+$ + CO $\rightarrow$ He+ O + C$^+$. Unlike the HCO$^+$ pathway, this reaction is irreversible. C$^+$ does not reform CO, but instead initiates hydrocarbon chemistry, producing species such as HCN and C$_2$H$_3$. These products have binding energies $\sim3$ to $6$ times higher than CO, so at $\sim$25--30~K they freeze onto grain, permanently sequestering the carbon. Over timescales of $\sim$10$^5$--10$^6$~years, this slow but steady process amounts to measurable depletion. The feature appears most prominently at high G$_0$ because gas-phase CO is abundant both above and below the affected layer. At low G$_0$, the disk midplane is cold enough that CO remains frozen as ice, so the depleted transition layer blends into the already CO-poor midplane.

\subsection{CO abundances across the full parameter space}

Having established in our fiducial models that increasing external FUV field strength enhances thermal desorption of CO ice through indirect heating, we now investigate whether these trends persist across the full range of physical parameters explored in our model grid. This broader analysis allows us to identify the conditions under which external irradiation significantly impacts CO chemistry, and to understand how stellar mass, disk mass, and radial location modulate this effect.

Our analysis focuses primarily on CO abundances in the disk interior, specifically at the midplane and below $z/r \lesssim 0.1$, for two reasons. First, this region is most relevant for planet formation, as it represents the reservoir from which planets accrete material. Second, abundances in the transition region and photoevaporative wind exhibit considerably more complexity and variability, reflecting the extreme and rapidly varying physical conditions in these regions. While these abundances are scientifically interesting, a detailed investigation of their behaviour is beyond the scope of this initial presentation of the parametric framework and is deferred to future studies.

\subsubsection{Vertical abundance profiles: varying disk mass}

\begin{figure*}
\centering
\includegraphics[clip=,width=1.0\linewidth]{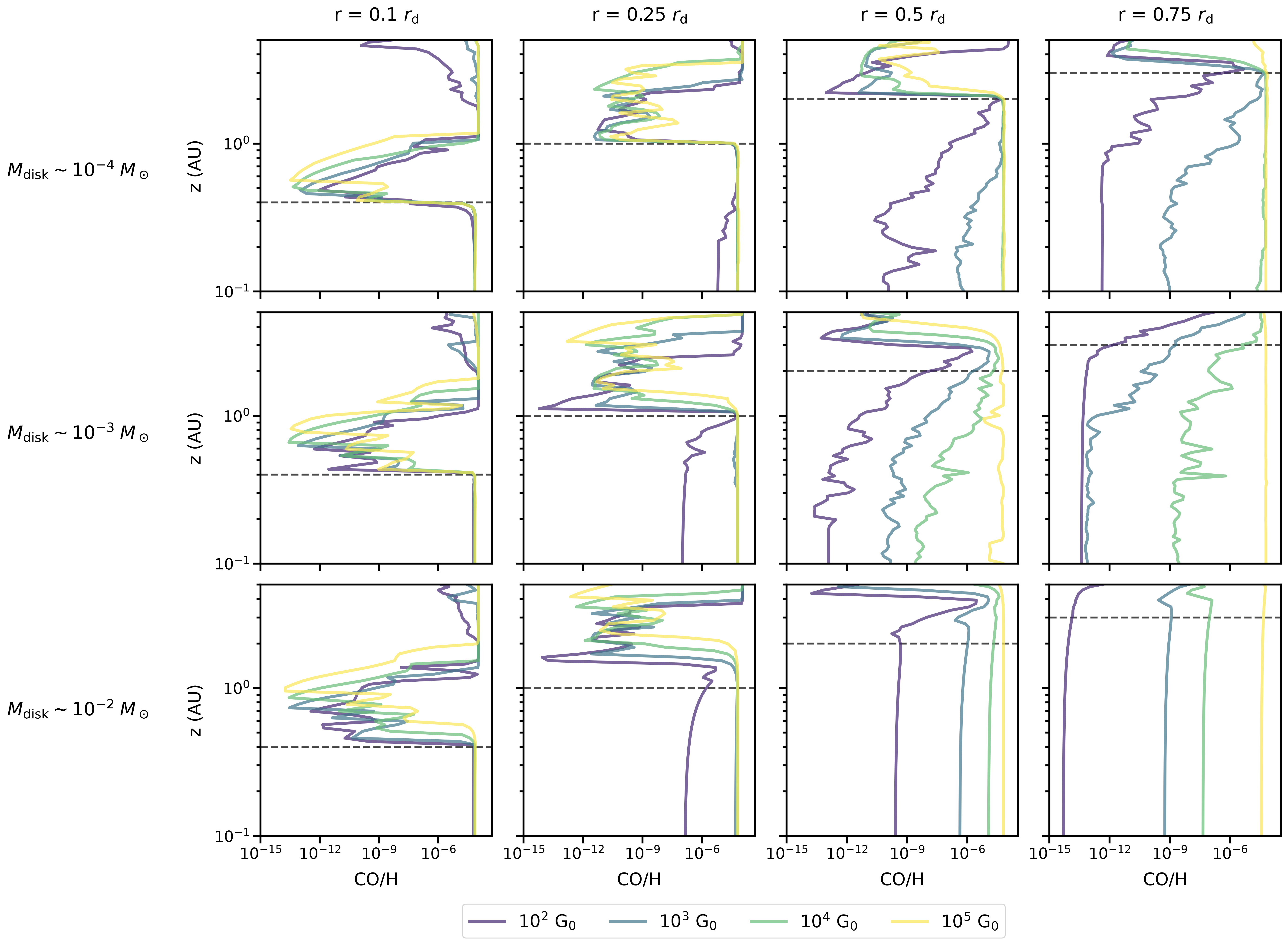}
\caption{Vertical CO abundance profiles at four radial locations (columns), for models with varying disk mass (rows). The $x$-axis shows the CO abundance relative to hydrogen (CO/H), while the $y$-axis shows height $z$ above the midplane in au. Different colored solid lines represent models with varying external FUV field strengths from 100 to 100,000 G$_0$. All models use the fixed fiducial parameters $r_\text{d}=40$ au and $M_*=0.6M_\odot$. Our analysis focuses on abundances below $z/r = 0.1$, denoted by the dashed horizontal lines.}
\label{fig_vertical_co_abundance}
\end{figure*}

We begin by examining how disk mass influences the vertical distribution of CO. Figure~\ref{fig_vertical_co_abundance} presents vertical abundance profiles extracted from models spanning disk masses of $M_\text{disk} \sim 10^{-4}$ to $10^{-2}$ $M_{\odot}$ (corresponding to $\Sigma_{1\mathrm{au}} = 10$ to $1000$ g cm$^{-2}$), while holding stellar mass ($M_* = 0.6$ $M_{\odot}$) and disk radius ($r_{\mathrm{d}} = 40$ au) fixed at their fiducial values. The profiles are extracted at four radial locations ($0.1$, $0.25$, $0.5$, and $0.75$ $r_{\mathrm{d}}$), providing representative sampling from the warm inner disk to the cold outer regions where CO freeze-out dominates under typical conditions.

In the innermost region ($r = 0.1$ $r_{\mathrm{d}}$), gas-phase CO is highly abundant, maintaining values close to CO/H $\sim 10^{-4}$  at the midplane across all models, and vertically up to $z/r \sim 0.1$. This behaviour does not vary with external FUV field strength, indicating that stellar FUV emission completely dominates the radiation environment at these small radii. The external field therefore has no discernible impact on either the vertical temperature structure or chemical evolution in the inner disk, regardless of disk mass.

Moving outward to $r = 0.25$ $r_{\mathrm{d}}$, the first signatures of the impact of external FUV become clear. At the lowest external field strength ($100$ G$_0$), CO becomes depleted from the gas phase at heights $z/r \lesssim 0.1$, reflecting freeze-out as temperatures drop below the CO sublimation point at this distance from the star. The degree of depletion scales strongly with disk mass. Low-mass disks ($M_\text{disk} \sim 10^{-4} \; M_\odot$) show modest depletion to CO/H $\sim 10^{-5}$ (one order of magnitude below the maximum), while high-mass disks ($M_\text{disk} \sim 10^{-2} \; M_\odot$) exhibit severe depletion to CO/H $\sim 10^{-7}$ (three orders of magnitude below maximum). This trend arises because more massive disks provide stronger attenuation of both stellar and external radiation, maintaining colder midplane temperatures that lead to higher freezeout rates.

However, increasing the external FUV field to $\geq 1000$ G$_0$ restores gas-phase CO abundances to near-maximum values across all disk masses. This recovery occurs because the midplane temperature at this location hovers close to the CO freeze-out threshold under nominal conditions. Most CO is frozen out, but only a modest temperature increase (a few Kelvin) suffices to thermally desorb the bulk of the ice reservoir and return CO to the gas phase.

At $r = 0.5$ $r_{\mathrm{d}}$, the trends become more pronounced and the dependence on both external FUV field strength and disk mass grows stronger. For low-mass disks under weak external irradiation ($100$ G$_0$), CO shows significant depletion at $z/r \lesssim 0.1$, with abundances dropping to CO/H $\sim 10^{-10}$. As external FUV increases to $10^3$ G$_0$, the abundance rises to $\sim 5 \times 10^{-7}$, and further increases in field strength fully restore CO to its maximum value, indicating almost complete thermal desorption of the CO ice reservoir.

For intermediate-mass disks, the pattern is similar but more extreme: depletion is more severe at all FUV field strengths, and only the highest external fields ($10^5$ G$_0$) succeed in returning gas-phase CO to maximum abundance. Interestingly, the highest-mass disks show less severe depletion than intermediate-mass disks when comparing identical external FUV field strengths. This seemingly counter-intuitive behaviour likely reflects the complex radiative transfer processes by which external FUV is converted to infrared radiation that heats the disk interior. While higher disk masses do provide stronger attenuation, they also drive higher photoevaporative mass-loss rates, producing denser winds. These denser winds more efficiently reprocess external FUV into infrared, which penetrates into the disk and warms the interior, partially compensating for the increased attenuation.

The outermost location ($r = 0.75$ $r_{\mathrm{d}}$) shows the strongest sensitivity to external FUV irradiation, as expected given that this cold region normally harbours the largest CO ice reservoir and receives minimal stellar radiation. For the lowest-mass disks, gas-phase CO at $z/r \lesssim 0.1$ is negligible under weak external fields ($100$ G$_0$), increases modestly to $\sim 10^{-9}$ at $1000$ G$_0$, and only returns to maximum abundance at the highest FUV field strengths. As disk mass increases, even more extreme external irradiation is required to thermally liberate the CO ice: the highest-mass disks show negligible gas-phase CO at $100$ G$_0$, remaining heavily depleted until the external field reaches $10^5$ G$_0$.

\begin{figure*}
\centering
\includegraphics[clip=,width=1.0\linewidth]{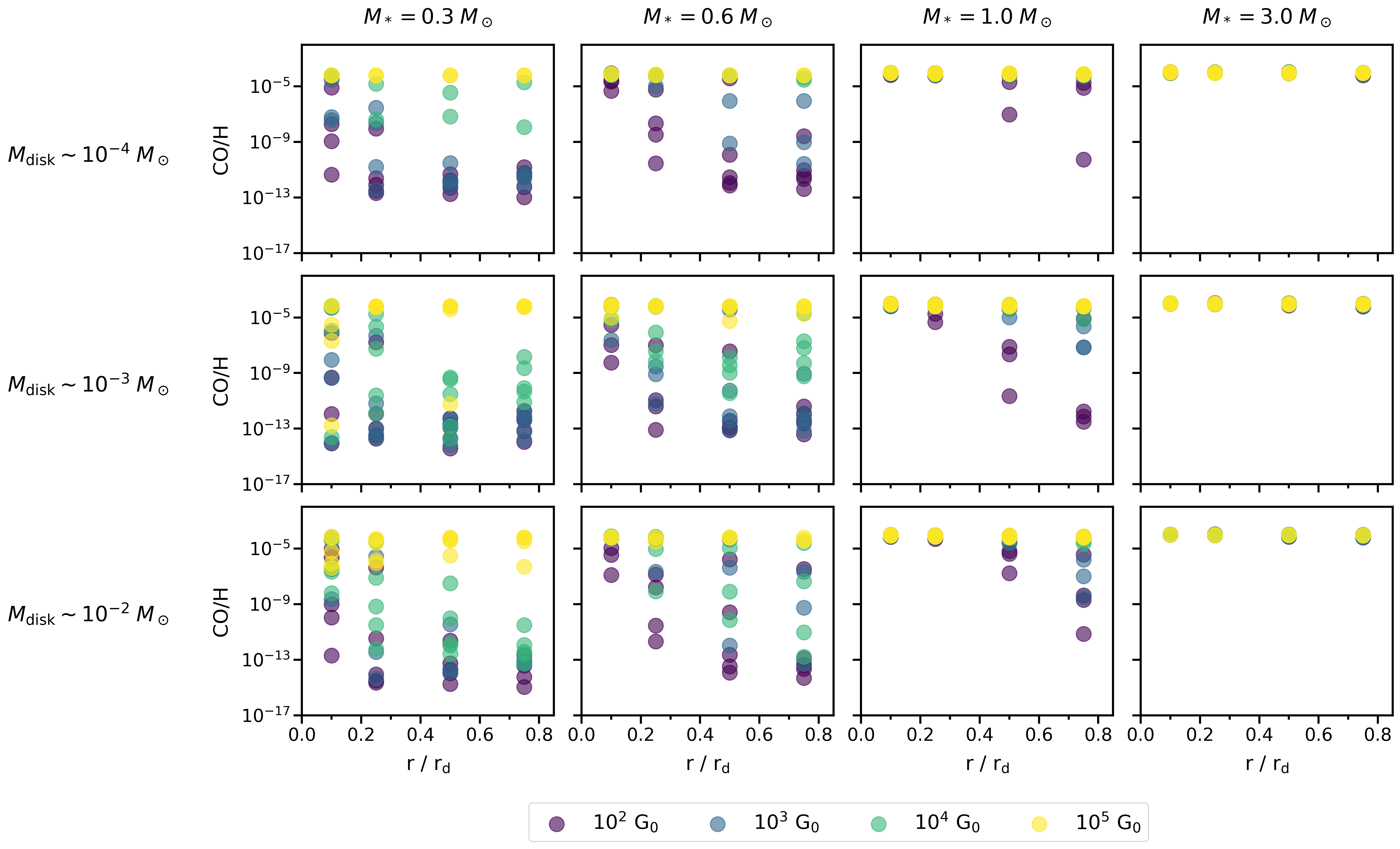}
\caption{Midplane CO abundance (CO/H) as a function of normalised disk radius ($r/r_\mathrm{d}$, where $r_\mathrm{d}$ is the disk outer edge) across the full parameter space. Rows show different disk masses ($M_\text{disk} \sim 10^{-4}$ to $10^{-2} \; M_\odot$) and columns show different stellar masses ($M_* = 0.3$ to $3.0\,M_\odot$). Point colors indicate external FUV field strength.}
\label{fig_midplane_CO}
\end{figure*}

\subsubsection{Midplane abundances: the full parameter space}

Figure \ref{fig_midplane_CO} extends our analysis across the complete parameter grid, showing midplane CO abundance as a function of normalized disk radius ($r/r_{\rm d}$) for all combinations of stellar mass (0.3 to 3.0 $M_\odot$), disk mass ($M_\text{disk} \sim 10^{-4}$ to $10^{-1} \; M_\odot$), and external FUV field strength ($10^2$ to $10^5$ G$_0$). This comprehensive view reveals some systematic trends in how external irradiation influences CO chemistry across a diverse range of disk and stellar properties.

For the most massive stars in our grid ($M_* = 3.0~M_\odot$), CO maintains near-maximum gas-phase abundances (CO/H $\approx 10^{-4}$) throughout the disk at all radii, regardless of external FUV field strength or disk mass. This behaviour reflects the dominant role of stellar heating, which is able to maintain midplane temperatures well above the CO sublimation temperature throughout the entire disk. Under these conditions, thermal desorption efficiently prevents CO freeze-out, and external FUV radiation has essentially no impact on the midplane CO distribution.

As stellar mass decreases to $M_* = 1.0~M_\odot$, external FUV irradiation begins to influence CO chemistry, though the effect remains confined primarily to the outer disk. In the inner regions ($r \lesssim 0.25~r_{\rm d}$), stellar heating continues to dominate, maintaining gas-phase CO abundances near maximum values across all external field strengths. Moving to intermediate radii ($r \approx 0.5~r_{\rm d}$) shows the first clear signatures of external FUV dependence. At the lowest external field strength (100 G$_0$), midplane temperatures cool sufficiently to drive significant CO depletion, with abundances dropping to CO/H $\sim 10^{-7}$ or lower. However, increasing the external FUV field to $\geq 10^3$ G$_0$ restores gas-phase CO to near-maximum values. In the outermost disk regions ($r \gtrsim 0.75~r_{\rm d}$), the dependence on external FUV becomes more pronounced. Under weak external irradiation, CO abundances become severely depleted across all disk masses. Progressively stronger external fields are required to restore gas-phase CO, with the threshold field strength increasing systematically with disk mass. High-mass disks ($\Sigma_{\rm 1au} = 1000$ g cm$^{-2}$) require external fields approaching $10^5$ G$_0$ to achieve near-complete thermal desorption of the CO ice reservoir in these cold outer regions.

The influence of external FUV extends to progressively smaller radii for lower-mass stars. At $M_* = 0.6~M_\odot$, even the inner disk ($r \approx 0.1~r_{\rm d}$) shows evidence of CO depletion under weak external irradiation (100 G$_0$), with abundances typically dropping by one to two orders of magnitude below maximum. Increasing external FUV fields restore gas-phase CO efficiently at this location, with fields $\geq 10^3$ G$_0$ sufficient to return abundances to near-maximum values across all disk masses. By $r \approx 0.25~r_{\rm d}$, CO depletion becomes significant at low external FUV across all disk masses, with abundances falling several orders of magnitude below maximum. The degree of depletion and the external FUV threshold required for thermal desorption both scale strongly with disk mass. Low-mass disks show moderate depletion and recover at intermediate external fields ($10^3$--$10^4$ G$_0$), while high-mass disks exhibit more severe depletion and require the strongest external fields ($\geq 10^4$ G$_0$) to restore gas-phase CO. In the outer disk ($r \gtrsim 0.5~r_{\rm d}$), CO becomes almost entirely frozen out at low external FUV strengths, with negligible gas-phase abundances (CO/H $\lesssim 10^{-12}$) across all disk masses. Only the most extreme external irradiation ($10^5$ G$_0$) succeeds in thermally liberating the bulk of the CO ice reservoir.

For the lowest stellar masses in our grid ($M_* = 0.3~M_\odot$), these trends intensify further. Significant CO depletion extends to even smaller disk radii under low and intermediate external FUV conditions. Notably, for higher mass disks, there are multiple instances at all radii where the gas-phase CO abundance remains depleted by many orders of magnitude even when the external FUV field reaches $10^5$ G$_0$. This indicates that for the coldest, most massive disk configurations, the indirect heating mechanism becomes insufficient to thermally desorb the CO ice reservoir, even under extreme external irradiation.

These systematic trends demonstrate that the impact of external FUV on midplane CO chemistry depends critically on the interplay between stellar heating, disk mass, radial location, and the strength of the external FUV field. External irradiation becomes increasingly important for lower-mass stars, higher-mass disks, and larger radii. These are precisely the conditions where stellar heating is weakest and CO freeze-out most prevalent under nominal conditions. The results highlight that external FUV fields $\geq 10^3$--$10^4$ G$_0$, which are typical of clustered star-forming regions like the Orion Nebula Cluster \citep[e.g.][]{storzer_hollenbach_1999}, can dramatically alter the CO gas/ice balance throughout much of the disk, with direct implications for the volatile inventory available to forming planets.

\subsection{Quantifying the influence of external FUV}

We have determined under which conditions external radiation affects the midplane  CO abundances. However, the effects depend on a complex interplay between stellar mass, disk mass, external field strength, and radial location. To better understand the relative importance of each parameter in determining midplane CO chemistry, we applied a Random Forest regression analysis to our complete model grid.

Random forests provide a robust non-parametric approach to identifying the key drivers of variance in complex datasets \citep{brieman_random_forest_2001}. The method constructs an ensemble of decision trees, each trained on a random subset of the data, and averages their predictions to reduce overfitting. Importantly for our purposes, the algorithm quantifies feature importance by measuring how much each parameter contributes to reducing prediction error across the ensemble.

We trained the model using the \texttt{RandomForestRegressor} implementation from \texttt{scikit-learn} \citep{pedregosa_scikitlearn_2011}, with 2000 estimators and unrestricted tree depth. The input features were stellar mass ($M_*$), disk surface density normalisation ($\Sigma_{1\mathrm{au}}$), disk radius ($r_\text{d}$), external FUV field strength ($F_{\mathrm{FUV}}$), and normalised radial position ($r/r_\text{d}$). The target variable was the midplane CO abundance (CO/H). We randomly partitioned the dataset into 80\% training and 20\% testing subsets.

\begin{figure}
\centering
\includegraphics[clip=,width=1.0\linewidth]{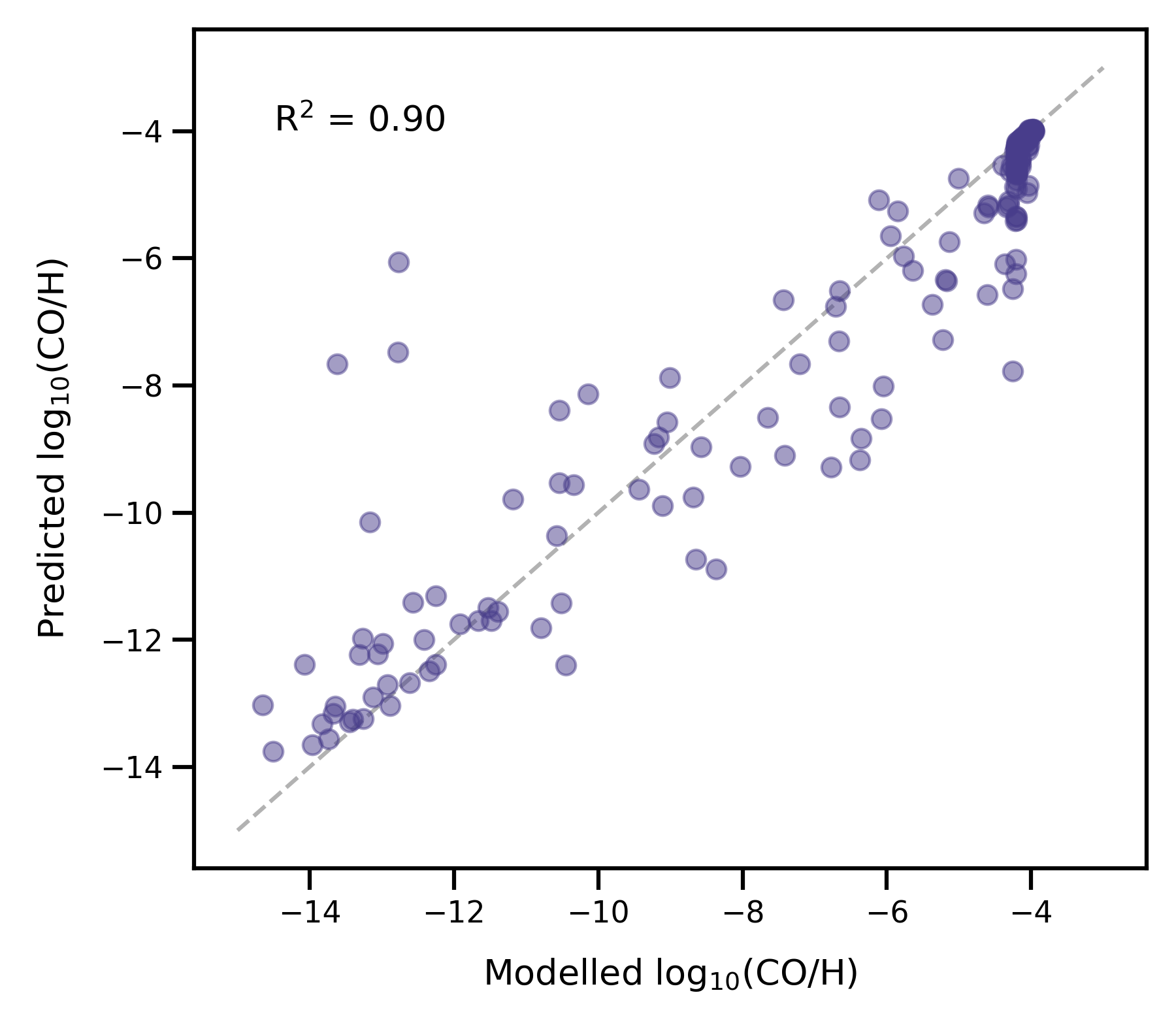}
\caption{Performance of the Random Forest regression model predicting midplane CO abundances. Each point represents a model from the independent test set, comparing predicted versus true CO/H values. The model achieves $R^2 = 0.90$, demonstrating that our selected physical parameters capture the dominant processes controlling midplane CO chemistry.}
\label{fig_random_forest}
\end{figure}

The trained model reproduces the variation in midplane CO abundance with high fidelity, achieving $R^2 = 0.90$ on the independent test set (Figure \ref{fig_random_forest}). This strong performance confirms that the selected parameters capture the principal physical processes governing CO chemistry across our parameter space. The model exhibits no systematic bias, with residuals distributed symmetrically around zero across the full range of CO abundances.

The feature importance analysis (Table \ref{table:random_forest}) reveals that stellar mass exerts the strongest control on midplane CO abundances, accounting for $\sim 39$\% of the explained variance. This dominance reflects the central role of stellar luminosity in establishing the disk's thermal structure. Higher-mass stars produce warmer disks, maintaining temperatures above the CO sublimation point across larger radial extents and thereby preventing freeze-out independently of external conditions.

The external FUV field strength is the second most important parameter, contributing $\sim$25\% to the explained variance. This substantial contribution reflects the efficiency of the indirect heating mechanism: external FUV radiation absorbed in the wind and disk atmosphere is reprocessed into infrared photons that penetrate deeply into the disk interior, raising midplane temperatures sufficiently to drive thermal desorption of CO ice across a wide range of disk configurations. The normalised radial position also contributes significantly ($\sim$20\%), consistent with the expected strong radial temperature gradients that control freezeout and thermal desorption rates. Disk size and mass play more modest roles (15\% and 5\%, respectively), acting primarily to regulate the efficiency with which both stellar and external radiation couple to the disk interior through their influence on optical depth.

These results quantify what our earlier analysis revealed: external FUV irradiation can fundamentally alter the CO gas-ice balance, with effects comparable in magnitude to variations in stellar heating. The relative importance of external versus stellar heating depends sensitively on stellar mass. For low-mass stars ($M_* \lesssim 0.6\,M_\odot$), where intrinsic stellar heating is weak, external FUV can dominate even at relatively small radii ($r/r_\text{d} \sim 0.1$). For massive stars ($M_* \gtrsim 3\,M_\odot$), stellar heating dominates, rendering the disk largely insensitive to its external radiation environment.

\begin{table}
\caption{Feature importance values from the Random Forest regression analysis. Values represent the fraction of explained variance attributed to each parameter in predicting midplane CO abundances.}        
\label{table:random_forest}      
\centering
\begin{tabular}{l l l}     
\hline\hline       
Feature & Description & Importance \\ 
\hline                    
$M_*$ & Stellar mass & 0.395 \\
$F_\text{FUV}$ & Integrated FUV field strength & 0.256 \\
$r / r_\text{d}$ & Scaled radius & 0.188 \\
$r_\text{d}$ & Disk outer edge & 0.116 \\
$\Sigma_\text{1au}$ & Surface density at $r=1$ au & 0.045 \\
\hline                  
\end{tabular}
\end{table}

\section{Discussion}
\label{sec:discussion}

\subsection{Model limitations and guidance}

The parametric model presented in this paper enables rapid generation of disk-wind density structures across a wide parameter space, but understanding both its capabilities and limitations is essential for proper application. Most fundamentally, this approach is not intended to replace full radiation-hydrodynamic simulations. Rather, it provides a computationally efficient framework for generating physically motivated density structures that capture the essential morphological features of externally irradiated systems. While these structures serve as effective inputs to thermochemical models and enable systematic parameter surveys that would be impractical with full simulations, they remain fundamentally approximate representations. Users requiring detailed studies of disk-wind dynamics or fine-scale structure should employ full hydrodynamical calculations.

The model has been extensively verified against hydrodynamical simulations spanning stellar masses from 0.3 to 3.0 $M_\odot$, disk radii from 20 to 150 au, surface density normalisations from 10$^1$ to 10$^4$ g cm$^{-2}$, and external FUV fields from $10^2$ to $10^5$ G$_0$. Within this range, the model performs reliably for the vast majority of physically meaningful parameter combinations. However, we have identified three specific conditions under which the parametric approach produces unphysical or unreliable structures (Section \ref{subsubsec:validating_2d_model}), and we caution against extrapolating beyond these validated ranges without careful testing.

Our framework focuses exclusively on FUV-driven external photoevaporation and does not account for several other potentially important physical processes. The model neglects external extreme ultraviolet (EUV) radiation, which can result in an ionisation front at some point in the flow \citep[i.e. giving the teardrop morphology of the proplyds][]{odell_wen_1994}. Even for many proplyds the ionisation front is quite far from the disk and so for the majority of the FUV field parameter space considered here, which extends up to the weaker UV fields that proplyds are exposed to, we expect this to not have a significant impact. However for some proplyds the ionisation front can be close to the disk \citep[][Amiot et al. submitted]{johnstone_1998} and there the impact would be more significant. Additionally, the model does not include \emph{internal} photoevaporation driven by high energy radiation from the central star itself. This may be particularly important for systems with weak external FUV fields where the contribution from the host star dominates, although in this case the impact of the disk density structure is only likely to be significant in the very inner few au. An additional limitation is that the external irradiation calculations assume an isotropic external radiation field, when in reality it could be more directional. There are not currently sufficient radiation hydrodynamic simulations \citep[only][which is only for the case of face-on irradiation]{richling_yorke_2000} to extend our parametric model to account for this. However this will be an important development in future. Users studying systems where these processes may be significant should interpret results cautiously or seek alternative modelling approaches that incorporate these mechanisms.

Given these limitations in the physics captured by the model, proper usage of its outputs is important. The temperature profiles and radiation fields calculated during model construction require particular care. These quantities exist purely to establish the density structure through our hydrostatic equilibrium solver and are deliberately simplified. Users must never rely on them directly for chemical modelling or producing synthetic observations. Instead, the resulting density structures should serve as inputs to dedicated radiative transfer codes that compute physically self-consistent temperature and radiation fields, as we demonstrated in Section \ref{sec:chemical_modelling} using \textsc{dali}. When performing such calculations, users should ensure that the external radiation field applied in the radiative transfer matches that specified during the parametric model construction to maintain consistency.

Beyond the primary physical inputs of stellar mass, disk mass, disk radius, and external FUV field strength, several parameters control specific aspects of the model behaviour. We provide calibrated prescriptions based on our extensive validation (Sections \ref{subsec:validating_1D} and \ref{subsubsec:validating_2d_model}), which perform well for typical applications. However, users may find it valuable to adjust these parameters when matching particular observations or incorporating into broader fitting frameworks. For example, the parameter $k$ governs the steepness of the vertical temperature gradient within the disk and thereby influences the vertical extent of the density structure computed through hydrodynamic equilibrium. The parameter $\gamma$ determines how rapidly the density transitions from the disk surface to the spherically diverging wind, effectively setting the spatial extent and morphology of the transition region. Our default formulation ties $\gamma$ to the stellar mass, external FUV field strength and disk size through calibrated scaling relations. However, users may prefer to treat $\gamma$ as a free parameter that can be adjusted manually or varied systematically, particularly when fitting to observations or as part of a broader Bayesian fitting framework. Complete documentation of all model parameters, their physical interpretations, and their effects on the resulting density structure is available in the online resources accompanying the \textsc{puffin} package at \texttt{puffin.readthedocs.io}

\subsection{Comparison with previous studies}

The parameter survey presented in Section \ref{sec:chemical_modelling} offers a framework for interpreting the diverse and sometimes contradictory findings reported in earlier observational and theoretical studies of externally irradiated disks. By examining how midplane CO abundances respond to variations in stellar mass, disk mass, radial location, and external FUV field strength, our results clarify several of the physical conditions that govern the chemical impact of external irradiation and may help to explain the apparent tension within the existing literature.

A central result from our modelling is that thermal desorption drives the enhancement of gas-phase CO at the disk midplane. This occurs through the indirect heating mechanism identified in \citet{keyte_haworth_2025}, in which external FUV radiation absorbed by the photoevaporative wind is reprocessed into infrared emission that penetrates deeply into the disk interior. Whether this mechanism produces observable chemical differences therefore depends on the presence of a substantial CO ice reservoir. Disks around higher-mass stars are already sufficiently warm to maintain CO in the gas phase even without external irradiation. As stellar mass decreases, external FUV begins to influence the midplane thermal structure. For $1\;M_\odot$ stars, this effect becomes apparent in the outer disk regions. For stars below $\lesssim 0.6 \; M_\odot$, external irradiation can substantially alter gas-phase CO abundances at the midplane across much larger radial extents. This dependence on both stellar mass and radial location may explain why some irradiated disks appear chemically similar to isolated systems while others show strong differences.

Recent observational work supports these predictions. \citet{boyden_eisner_2023}, for example, reported near-ISM CO abundances in a sample of massive Orion disks predominantly hosted by sub-solar-mass stars. These disks likely retain substantial CO ice reservoirs in their outer regions, making them especially susceptible to the enhanced thermal desorption produced under strong external fields. Although the limited spatial resolution of the observations raises the possibility that part of the measured flux originates in a wind rather than the bound disk, the overall trend toward near-ISM abundances is consistent with the expectation that external heating reduces the efficiency of CO freeze-out. Similarly, \citet{goicoechea_2024} infer an undepleted carbon reservoir in the irradiated disk d203-506, which also orbits a low-mass host star ($M_* \sim 0.3\,M_\odot$) and is exposed to an intense external FUV field of $\sim 10^4$ G$_0$. Both results agree with our models, which suggest that field strengths of $\gtrsim 10^4$ G$_0$ are required to significantly enhance gas-phase CO abundances across the entire disk midplane.

In contrast, other observations suggest the impact of external irradiation may be more subdued. ALMA studies of disks exposed to relatively weak external radiation fields ($\lesssim 10^3$ G$_0$) find compositions indistinguishable from those of isolated disks \citep{diaz_berrios_2024}. Our models likewise show that moderate irradiation generally cannot overcome stellar heating or significantly affect CO freeze-out, even at large radii where the stellar FUV field becomes less dominant. JWST spectra of the irradiated systems XUE 1 \citep{ramirez_tannus_2023} and d203-504 \citep{schroetter_2025} also show inner-disk CO abundances comparable to those in non-irradiated disks, despite exposure to external FUV fields of $\sim10^5$ G$_0$ and $10^4$ G$_0$, respectively. Because JWST probes only the very innermost few au, where stellar heating dominates, such a result is consistent with our finding that even strong external FUV fields influence the midplane primarily where intrinsic temperatures are low enough for substantial CO ice to exist.

Theoretical studies further reflect this diversity. Models with moderate external fields ($1-100$ G$_0$) predict only minor CO enhancement restricted to the cold, low-density outer disk, where photodesorption can play a modest role \citep{gross_cleeves_2025}. Our results reproduce this behaviour, showing that stellar heating already maintains most CO in the gas phase inside roughly half the disk radius for solar-mass stars, leaving only the outermost regions sensitive to such mild photodesorption effects. At higher irradiation levels, our findings agree with earlier simulations that showed CO remaining in the gas phase across the entire midplane once external fields exceed $\sim 10^5$ G$_0$ \citep{walsh_2013}. Similarly, \citet{calahan_2025} modelled disks subject to high external FUV fields (up to $10^6$ G$_0$), focusing particularly on the inner disk ($\lesssim 10$ au), finding that although CO is abundant under all irradiation conditions this close to the star, external fields of $\gtrsim 10^4$ G$_0$ dramatically increase the disk-integrated column density of CO that is observable at infrared wavelengths accessible to eg. JWST. While this result is not directly comparable to the model grid presented in this work, which focuses only on midplane CO abundances, the overall conclusion is similar: external fields $\gtrsim 10^4$ G$_0$ are generally required to produce significant enhancements in gas-phase CO across a broad range of disk and stellar conditions.

It is also important to recognise that our chemical models are computed on top of static density structures that do no capture all potentially physically relevant processes. For example, our models do not account for dynamical processes such as radial drift, or advection of gas and ice within the disk or photoevaporative flow. One-dimensional external photoevaporation models that couple dynamics with chemistry suggest that the chemical composition of the inner disk can remain essentially unchanged despite strong external irradiation, as the rapid inward drift of icy pebbles continuously replenishes volatile reservoirs and sets the inner disk composition \citep{ndugu_2024}. This mechanism could explain why some observations of inner disk chemistry in irradiated systems show compositions similar to isolated disks. Similarly, one-dimensional models that combine the dynamics and chemistry of photoevaporative flows demonstrate that the chemical composition of the wind can differ substantially from predictions based on static structures, with molecular ices surviving significantly further into the flow than equilibrium calculations suggest \citep{keyte_ran_2025}. These dynamical effects arise because the advection timescale throughout the flow can be shorter than the chemical timescales for sublimation, allowing ices to persist in regions where static models predict their complete destruction. Extending such coupled hydrodynamical and chemical calculations to two dimensions remains computationally expensive, which is why essentially all existing two-dimensional chemical models of externally irradiated disks compute chemistry on static density structures. Interpreting the results from such models should always consider this limitation and recognise that dynamical processes may modify the chemical distributions and abundances predicted from static calculations.

Taken together, the variety of outcomes reported in previous observational and modelling studies does not necessarily reflect contradictory chemical behaviour. Instead, it arises from the interplay of stellar mass, external FUV field strength, disk mass, and the specific radial regions probed by each analysis. By systematically mapping how midplane CO responds across this parameter range, our survey identifies the regimes in the influence of the external FUV field is greatest. Extending this type of analysis beyond CO to additional molecular species and to the full disk–wind structure will be essential for building a unified understanding of the chemistry of externally irradiated disks.

\subsection{Future applications}

The efficient parametric framework presented in this paper opens several promising avenues for future research that address fundamental questions about planet formation in clustered environments.

A pressing challenge for observational studies of externally irradiated disks is spatial resolution. Even the nearest externally irradiated systems in regions like the Orion Nebula Cluster ($\sim400$ pc; \citealt{jeffries_2007, kounkel_2017}) are typically resolved into only a few resolution elements with current ALMA observations \citep[e.g.][]{boyden_eisner_2023}, making it difficult to distinguish between emission originating from the disk interior versus the photoevaporative wind. In more distant irradiated regions such as the Carina Nebula (2.3 kpc), externally photoevaporating systems generally appear as unresolved point sources or are only marginally resolved \citep[e.g.][]{mesa_delgado_2016}. Developing observational diagnostics that remain informative without high spatial resolution is therefore critical. Our framework enables systematic exploration of molecular line ratios and integrated fluxes that could serve as robust tracers of external irradiation strength, complementing recent work on atomic line diagnostics in the UV to far-infrared \citep{peake_2025}. Previous studies have identified species such as CN, HCN, and HCO$^+$ as potentially sensitive to high external FUV fields \citep{walsh_2013, gross_cleeves_2025}, but comprehensive parameter studies are needed to establish which molecular tracers provide the most reliable diagnostics across diverse disk masses, stellar types, and irradiation conditions.

Such observational diagnostics are essential not only for characterising external irradiation, but also for probing how these environments affect the volatile reservoirs available for planet formation. External irradiation can fundamentally alter the thermal structure of protoplanetary disks through indirect heating, as demonstrated in Section \ref{sec:chemical_modelling}, enhancing thermal desorption of volatile ices with direct consequences for planet formation. The systematic effects of external irradiation on the locations of key snowlines, gas-phase C/O ratios, and the extent to which ice reservoirs may be depleted before incorporation into planetesimals remain poorly understood. While our analysis focused on CO, extending this framework to other critical volatiles such as H$_2$O, CO$_2$, complex organics, sulfur-bearing species, and nitriles is essential for understanding the full compositional inventory available during planet formation. These molecules are observable with JWST and crucial for planetary atmospheres, making such studies directly testable against observations. Systematic exploration of how volatile budgets vary with stellar mass, disk mass, radial location, and field strength can provide constraints for planet formation models and predictions for the compositional diversity of exoplanets formed in clustered versus isolated environments.

These compositional changes raise a fundamental question about the origins of disk chemistry: to what extent do protoplanetary disks inherit their chemical composition from earlier stages of star formation versus undergoing chemical reset during disk evolution? Recent modelling suggests that extreme external FUV fields ($\gtrsim 10^4$ G$_0$) can partially reset even inner disk chemistry \citep{calahan_2025}, though the specific conditions under which this occurs remain somewhat unconstrained. Population-level studies spanning the full range of irradiation conditions encountered in star-forming regions can establish quantitative relationships between external radiation environment and the degree of chemical processing, addressing whether planet-forming disks retain a chemical memory of their origins or whether external irradiation effectively resets their composition.

Answering these questions will require identifying the specific physical and chemical mechanisms operating in externally irradiated disks. The computational efficiency of our framework makes it practical to perform the large parameter studies needed to explore formation and destruction pathways for specific molecules that might serve as sensitive probes of FUV exposure. Recent observations show that emission from species such as OH, H$_3^+$, CH$^+$, and CH$_3^+$ can provide insights into specific physical and chemical processes at play in irradiated environments, acting as valuable tracers of FUV-driven chemistry \citep[e.g.][]{zannese_2024, zannese_2025, goicoechea_2025, schroetter_2025b}. Systematic parameter surveys using our modelling framework could help identify additional molecular species whose abundances and emission properties provide robust diagnostics of irradiation conditions, while simultaneously constraining the mechanisms responsible for the volatile depletion and compositional changes discussed above.

These applications represent only a subset of questions now within reach. By enabling efficient exploration of the complex parameter space governing externally irradiated disks, our parametric framework provides the foundation for understanding how environment shapes the chemical evolution of planet-forming disks and, consequently, the composition of planetary systems throughout the Galaxy.

\section{Conclusions}
\label{sec:conclusions}

We have developed \textsc{puffin}, a parametric framework for rapidly generating density structures of externally irradiated protoplanetary disks with photoevaporative winds. By enabling systematic exploration across parameter spaces where full radiation-hydrodynamic simulations are computationally prohibitive, this approach facilitates comprehensive studies of disk-wind chemistry. We demonstrate its utility by investigating how external FUV irradiation influences CO abundances across diverse stellar and disk properties. Our key findings are:

\begin{enumerate}
    \item The parametric model reproduces hydrodynamical simulations with typical factor-of-two accuracy across stellar masses of 0.3--3.0~$M_\odot$, disk radii of 20--150~au, and external FUV fields of $10^2$--$10^5$~$G_0$. Validation against 600+ 1D models from the \textsc{fried} grid and representative 2D simulations confirms the framework captures essential disk-wind physics across $\gtrsim 98$\% of the validated parameter space.
    \item \textsc{puffin} generates complete 2D density structures in seconds to minutes versus weeks to months for equivalent hydrodynamical calculations, enabling parameter surveys that were previously impractical. The framework is publicly available as a Python package with mass-loss rate automatically interpolation from the \textsc{fried} grid.
    \item Apply our parametric framework to a large grid of chemical models shows that external FUV substantially enhances midplane gas-phase CO abundances through indirect heating. UV photons absorbed in the photoevaporative wind reprocess into infrared radiation that penetrates deep into the disk interior, raising midplane temperatures above the CO sublimation point even where UV itself is completely attenuated at the disk surface.
    \item The effectiveness of external FUV in enhancing CO abundances depends critically on stellar mass, disk mass, and radial location. For massive stars ($M_* \gtrsim 3$~$M_\odot$), stellar heating dominates completely and external irradiation has negligible impact on midplane chemistry. For lower-mass stars ($M_* \lesssim 0.6$~$M_\odot$), external fields $\gtrsim 10^3$~G$_0$ can restore gas-phase CO across much of the disk. Higher-mass disks require stronger external fields ($\gtrsim 10^4$--$10^5$~G$_0$) for comparable desorption due to increased optical depth, though this is partially offset by denser winds that more efficiently reprocess FUV into penetrating infrared radiation. The effect strengthens systematically toward larger radii where stellar heating weakens and CO ice reservoirs are largest.
    \item Random Forest regression analysis quantifies that stellar mass contributes 39\% and external FUV field strength 26\% of explained variance in midplane CO abundances, with radial position, disk size, and disk mass contributing 19\%, 12\%, and 5\% respectively. The comparable importance of stellar mass and external FUV confirms that external irradiation is a first-order effect rather than a minor perturbation on disk chemistry.
\end{enumerate}

This work shows that protoplanetary disk chemistry cannot be fully understood without accounting for the radiation environment. Our computationally efficient parametric framework now enables systematic studies of how external irradiation influences volatile reservoirs, snowline locations, and molecular complexity across the diverse conditions of clustered star formation. Since most planets form in irradiated environments, these effects are crucial for interpreting both disk chemistry and planetary composition.

\section*{Acknowledgements}

LK is funded by UKRI guaranteed funding for a Horizon Europe ERC consolidator grant (EP/Y024710/1). TJH acknowledges UKRI guaranteed funding for a Horizon Europe ERC consolidator grant (EP/Y024710/1) and a Royal Society Dorothy Hodgkin Fellowship.

\section*{Data Availability}

The GitHub repository for \textsc{puffin} can be found at \url{https://github.com/lukekeyte/PUFFIN/}. The \textsc{fried} grid of hydrodynamical simulations are available at \url{https://github.com/thaworth-qmul/FRIEDgrid/}.



\bibliographystyle{mnras}
\bibliography{bibliography} 




\appendix

\newpage
\section{Stellar FUV approximation}
\label{appendix:fuv_fraction}

In our 2D model, we estimate the stellar FUV luminosity using the empirical scaling relations established by \citet{eker_2018}. These relations enable us to derive the stellar effective temperature, $T_\text{eff}$, directly from the stellar mass. Assuming the star emits as a blackbody, we then integrate the Planck function at this effective temperature over the FUV wavelength range (912–2000 \AA) to determine the fraction of the total stellar luminosity emitted in the FUV.

The \citet{eker_2018} relations are based on high-precision observations of 509 main-sequence stars and provide three interrelated functions: mass-luminosity (MLR), mass-radius (MRR), and mass-temperature (MTR) relations. These relations are calibrated over different mass ranges. For masses $M_* \leq 1.5 M_\odot$, the MRR is empirically determined while the effective temperature is computed from the MLR and MRR. For higher masses ($M_* > 1.5 M_\odot$), the MTR is empirically determined while the radius is computed from the MLR and MTR.

The six-piece mass-luminosity relation is:
\begin{equation}
    \log L_* = \begin{cases}
        2.028 \log M_* - 0.976 & \text{if } 0.179 M_\odot \leq M_* \leq 0.45 M_\odot \\
        4.572 \log M_* - 0.102 & \text{if } 0.45 M_\odot < M_* \leq 0.72 M_\odot \\
        5.743 \log M_* - 0.007 & \text{if } 0.72 M_\odot < M_* \leq 1.05 M_\odot \\
        4.329 \log M_* + 0.010 & \text{if } 1.05 M_\odot < M_* \leq 2.4 M_\odot \\
        3.967 \log M_* + 0.093 & \text{if } 2.4 M_\odot < M_* \leq 7.0 M_\odot \\
        2.865 \log M_* + 1.105 & \text{if } 7.0 M_\odot < M_* \leq 31.0 M_\odot \\
        \end{cases}
\end{equation}
The mass-radius relation for low-mass stars ($0.179 M_\odot < M_* \leq 1.5 M_\odot$) is:
\begin{equation}
    R_* = 0.438 \cdot M_*^{2} + 0.479 \cdot M_{*} + 0.075
\end{equation}
The mass-temperature relation for high-mass stars ($1.5 M_\odot < M_* \leq 31.0 M_\odot$) is:
\begin{equation}
    \log T_{\text{eff}} = -0.170 \cdot \log M_*^2 + 0.888 \cdot \log M_8 + 3.671
\end{equation}

\begin{figure}
\centering
\includegraphics[clip=,width=1\linewidth]{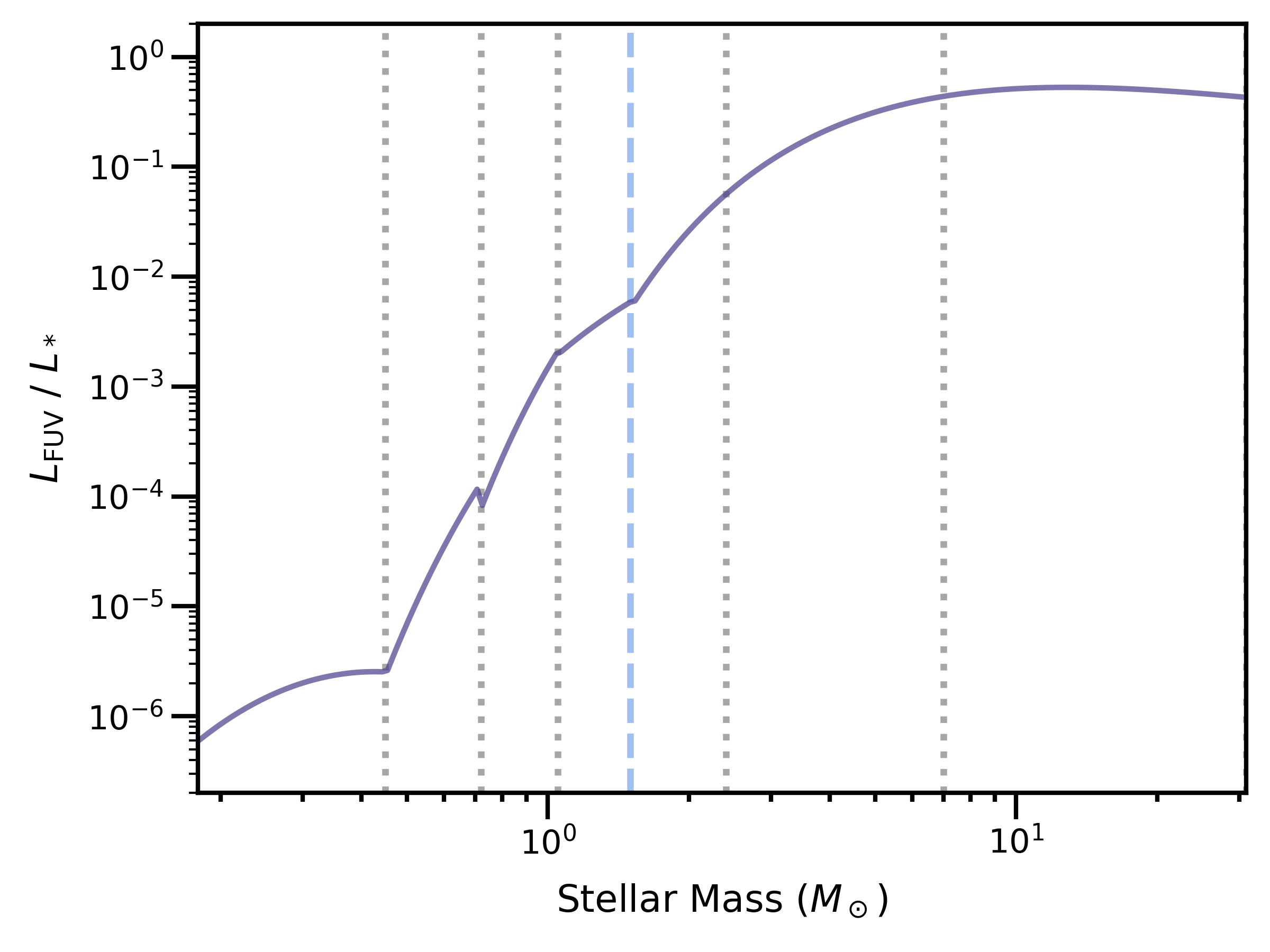}
\caption{FUV luminosity fraction, $L_\text{FUV}/L_*$, versus stellar mass derived using the \citet{eker_2018} stellar relations. Vertical lines indicate break points in the empirical mass-luminosity (dotted) and mass-radius/temperature (dashed) relations.}
\label{fig_fuv_fraction}
\end{figure}

Once the effective temperature is determined, we model each star as a blackbody and calculate the FUV luminosity fraction by integrating the Planck function over the wavelength range 912--2000~\AA:
\begin{equation}
    f_{\text{FUV}} = \frac{L_{\text{FUV}}}{L_*} = \frac{\int_{912}^{2000} B_\lambda(T_{\text{eff}}) \, d\lambda}{\int_0^\infty B_\lambda(T_{\text{eff}}) \, d\lambda}
\end{equation}
\noindent where $B_\lambda(T_{\text{eff}})$ is the Planck function. Figure \ref{fig_fuv_fraction} shows the resulting FUV fraction as a function of stellar mass across the range $0.179 M_\odot \leq M_* \leq 31.0 M_\odot$. For stellar masses outside this range, we extrapolate using the boundary values.

Although this treatment is highly simplified, relying on relationships derived for main sequence stars, it provides a reasonable first-order approximation for the stellar FUV fraction. In practice, the stellar FUV contribution generally has minimal impact on the resulting wind structure, since the wind is tapered close to the host star where the stellar FUV field is strongest.

\section{Examples of the 2D parametric model}
\label{appendix:examples_2d_parametric_model}

In this section we present representative examples of the 2D density structures generated by our parametric model, demonstrating its behaviour across a wide parameter space. While not exhaustive, these examples illustrate the general model morphology across a wide range of conditions.

\begin{figure*}
\centering
\includegraphics[clip=,width=1\linewidth]{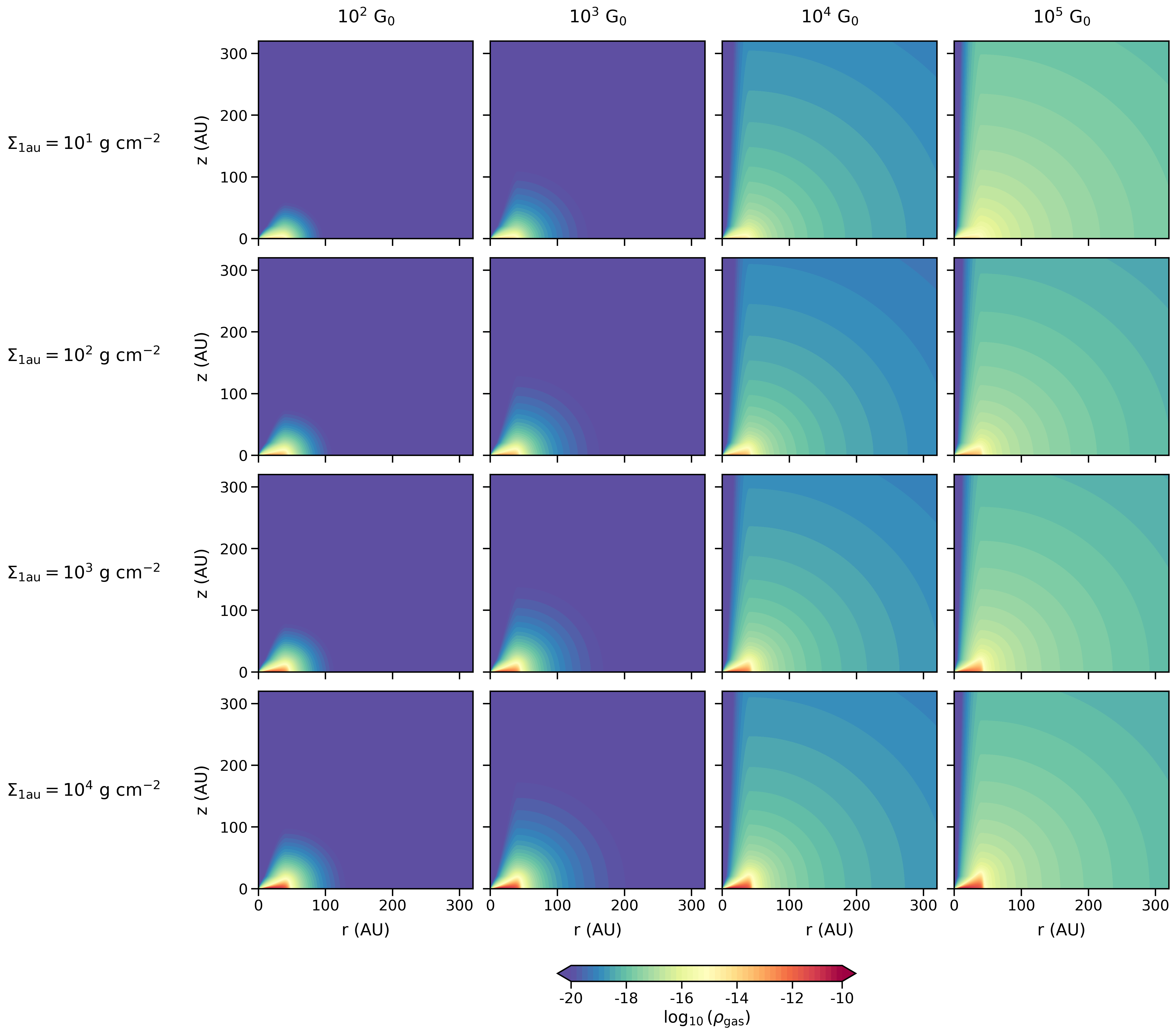}
\caption{Illustrative set of density structures generated with our 2D parametric model, shown as a function of external FUV field strength (columns) and disk mass (varied through the surface density normalisation, $\Sigma_\text{1au}$, rows). The disk radius is fixed at $r_\text{d} = 40$ au, and the stellar mass is fixed at $M_* = 1.0,M_\odot$.}
\label{fig_examples_sigma}
\end{figure*}

Figure \ref{fig_examples_sigma} shows model density structures for a range of external FUV field strengths and surface density normalisations, while the disk radius and stellar mass are kept fixed at $r_\text{d} = 40$ au and $M_* = 1.0,M_\odot$.

\begin{figure*}
\centering
\includegraphics[clip=,width=1\linewidth]{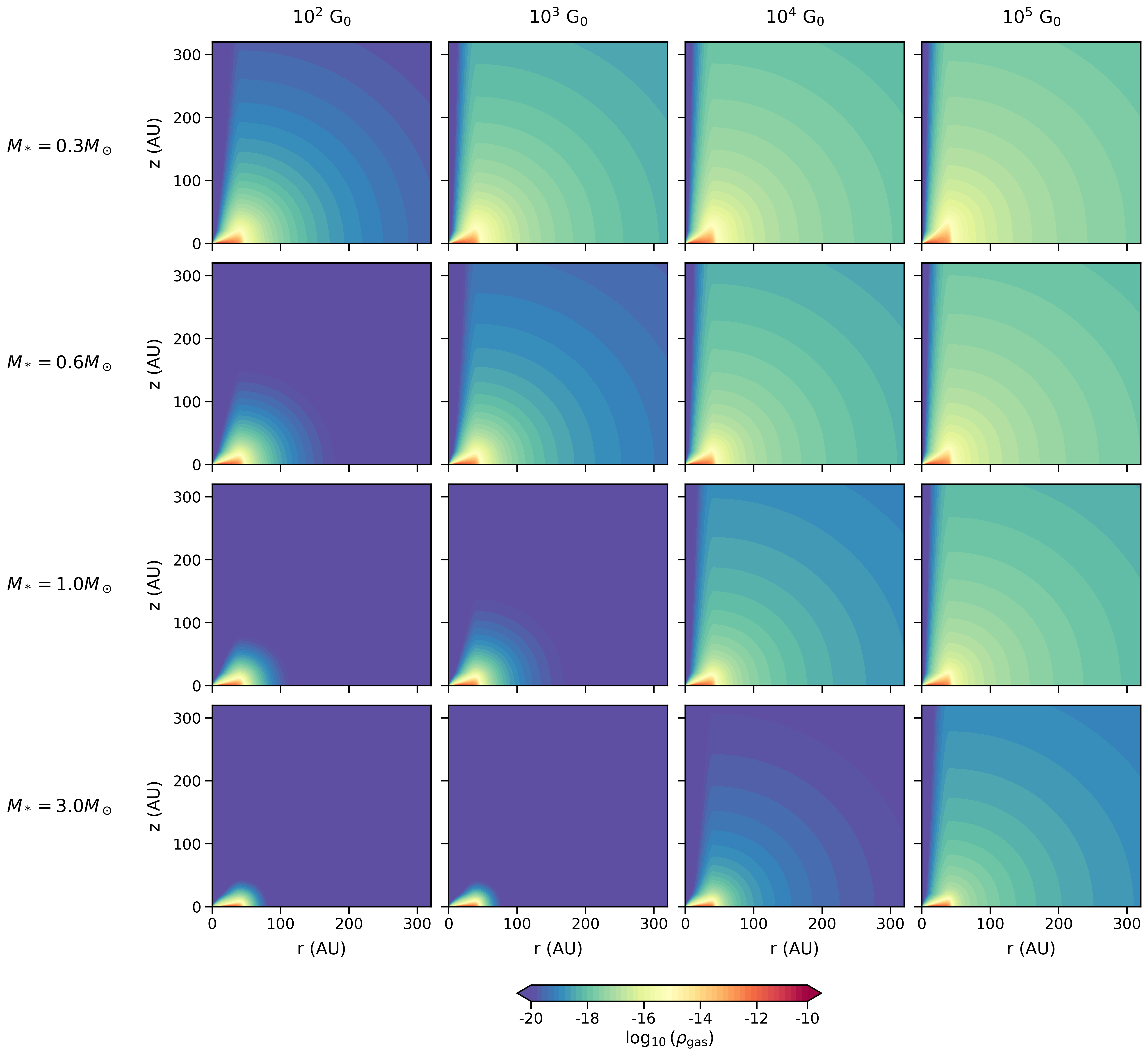}
\caption{Illustrative set of density structures generated with our 2D parametric model, shown as a function of external FUV field strength (columns) and stellar mass (rows). The disk radius is fixed at $r_\text{d} = 40$ au, and the surface density normalisation is fixed at $\Sigma_\text{1au} = 1000$ g cm$^{-2}$.}
\label{fig_examples_mstar}
\end{figure*}

Figure \ref{fig_examples_mstar} shows model density structures for a range of external FUV field strengths and stellar masses, while the disk radius and surface density normalisation are kept fixed at $r_\text{d} = 40$ au and $\Sigma_\text{1au} = 1000$ g cm$^{-2}$.

\begin{figure*}
\centering
\includegraphics[clip=,width=1\linewidth]{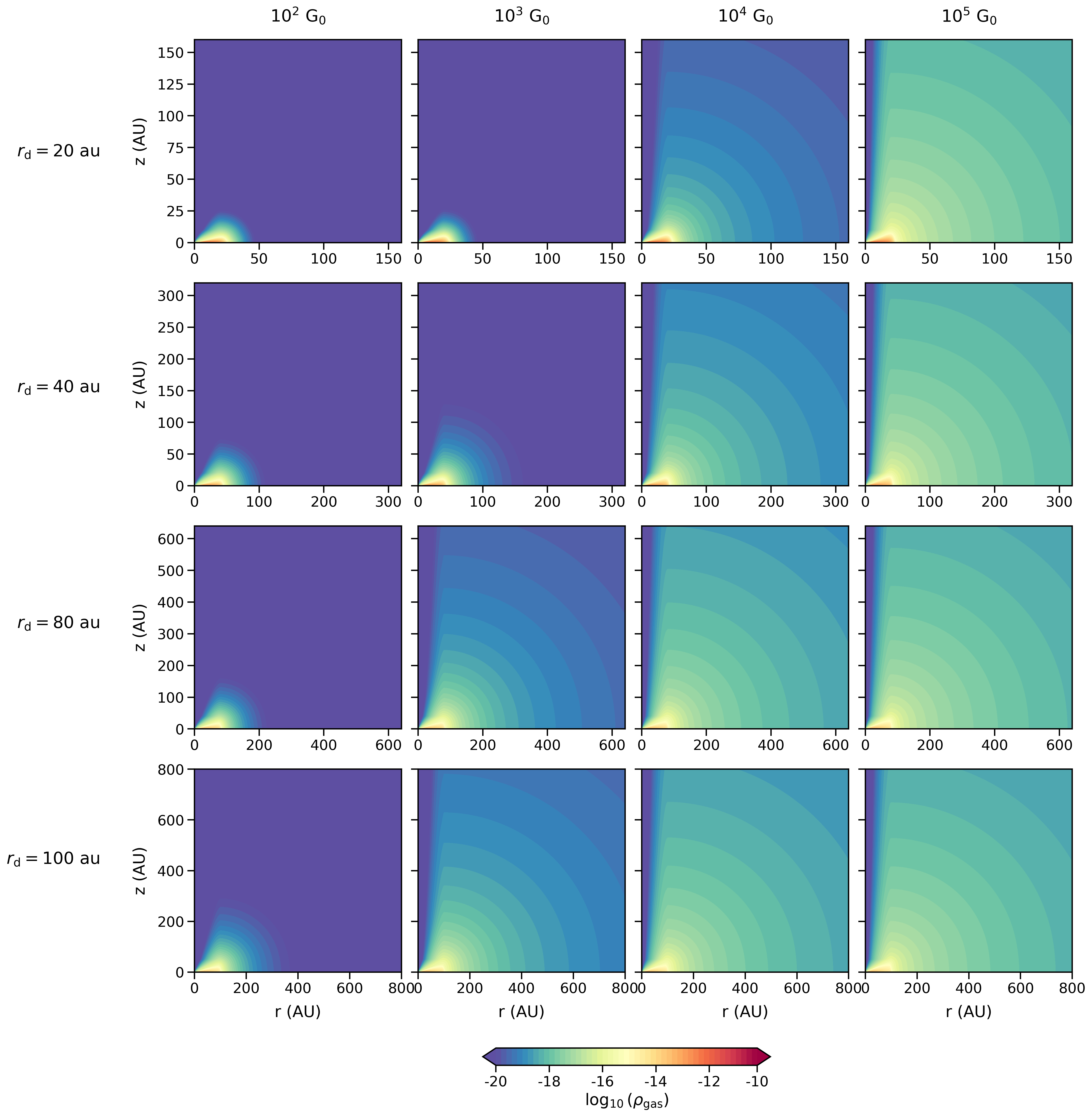}
\caption{Illustrative set of density structures generated with our 2D parametric model, shown as a function of external FUV field strength (columns) and disk size (rows). The stellar mass is fixed at $M_* = 1 M_\odot$, and the surface density normalisation is fixed at $\Sigma_\text{1au} = 1000$ g cm$^{-2}$.}
\label{fig_examples_rd}
\end{figure*}

Figure \ref{fig_examples_rd} shows model density structures for a range of external FUV field strengths and disk sizes, while the stellar mass and surface density normalisation are kept fixed at $M_* = 1 M_\odot$ and $\Sigma_\text{1au} = 1000$ g cm$^{-2}$.

\bsp	
\label{lastpage}
\end{document}